\documentclass[a4paper,12pt]{article}
\usepackage{jcappub}
\usepackage{amsthm,graphicx,multirow}
\usepackage{epsfig}
\usepackage{latexsym, amssymb} 
\usepackage{amsmath}
\def\beq{\begin{equation}}
\def\eeq{\end{equation}}
\def\br{\begin{eqnarray}}
\def\er{\end{eqnarray}}
\def\benu{\begin{enumerate}}
\def\efnu{\end{enumerate}}
\def\nn{\nonumber}

\def\l{\left}
\def\r{\right}

\def\cl{{\cal C}_{\ell}}

\begin{document}
\title{Ruling out the power-law form of the scalar primordial spectrum}
 \author[a]{Dhiraj Kumar Hazra} 
 \author[a,b]{Arman Shafieloo}
 \author[c,d,e]{George F. Smoot}
 \author[f]{Alexei A. Starobinsky}
 
\affiliation[a]{Asia Pacific Center for Theoretical Physics, Pohang, Gyeongbuk 790-784, Korea}
\affiliation[b]{Department of Physics, POSTECH, Pohang, Gyeongbuk 790-784, Korea}
\affiliation[c]{Lawrence Berkeley National Laboratory, Berkeley, CA 94720, USA}
\affiliation[d]{Institute for the Early Universe, Ewha Womans University Seoul, 120-750, Korea}
\affiliation[e]{Paris Centre for Cosmological Physics, APC, Universite Paris Diderot, France}
 \affiliation[f]{Landau Institute for Theoretical Physics RAS, Moscow, 119334, Russia}
\emailAdd{dhiraj@apctp.org, arman@apctp.org, gfsmoot@lbl.gov, alstar@landau.ac.ru}

\abstract 
{Combining Planck CMB temperature~\cite{Ade:2013kta} and BICEP2 B-mode polarization data~\cite{Ade:2014gua,Ade:2014xna} 
we show qualitatively that, 
assuming inflationary consistency relation, 
the power-law form of the scalar primordial spectrum is ruled out at more than $3\sigma$ CL. 
This is an important finding, since the power-law form of the scalar primordial spectrum 
is one of the main assumptions of concordance model of cosmology and also a direct prediction of many 
inflationary scenarios. We show that a break or step in the form of the primordial scalar perturbation spectrum, 
similar to what we studied recently analyzing Planck data~\cite{Hazra:2013nca}, can address both Planck and BICEP2 results simultaneously. 
Our findings also indicate that the data may require more flexibilities than what running of scalar spectral index can provide. 
Finally we show that an inflaton potential, originally appeared in~\cite{Bousso:2013uia}, can generate 
both the step and the break model of scalar primordial spectrum in two different limits. 
The discussed potential is found to be favored by Planck data but marginally disfavored by BICEP2 results
as it produces slightly lower amplitude of tensor primordial spectrum. 
Hence, if the tensor-to-scalar ratio ($r$) quoted by BICEP2 persists, it is of importance that we generate inflationary
models with large $r$ and at the same time provide suppression in scalar primordial spectrum at large scales.
}

\maketitle
\section{Introduction}

The primary goal of physical cosmology is to find an accurate model of the Universe. 
The current standard model of cosmology, also known as concordance model, is a spatially flat FLRW Universe consist of weakly interacting cold dark matter, 
cosmological constant and baryons all in the context of power-law form of the primordial perturbations. 
This simple power law form is a natural result if, for example, one assumes slow-roll Inflation.
While the features in the primordial power spectrum (PPS) have been subject of various studies, cosmological observations before BICEP2 B-mode polarization results~\cite{Ade:2014gua,Ade:2014xna} have been all 
essentially consistent to  the power-law form of the primordial perturbation spectrum. 
The very high tensor-to-scalar ratio from the BICEP2 B -Mode polarization  at large angular scales may change this picture. 

In fact it seems to be hard to assume a power-law form of the primordial spectrum and have a good fit to both Planck temperature~\cite{Ade:2013kta} (and WMAP~\cite{Hinshaw:2012fq} 
low-$\ell$ polarization) and BICEP2 B-Mode observations simultaneously. 
Assuming that BICEP2 has detected B-modes corresponding to a tensor-to-scalar ratio $r_{0.002}\sim0.2$ (defined at pivot scale $0.002{\rm Mpc^{-1}}$) indicates that the high tensor component will add power to large angular scale temperature anisotropy. 
On the other hand Planck data indicates a mild suppression
in large angular scale power compared to standard power law $\Lambda$CDM model~\cite{Ade:2013kta}. In fact there were hints of large scale suppression
in scalar power since the first year results of WMAP~\cite{Peiris:2003ff}. Several model independent reconstruction methods~\cite{Hazra:2013nca,reconstruction-all} using different datasets 
too hint towards suppression in scalar power.
This mild suppression limits the tensor primordial spectra to be higher than $r_{0.002}< 0.12$ (95\% C.L.)~\cite{Ade:2013uln} assuming a
primordial perturbation power law. 
In this paper we address the consistency of the power-law PPS with combination of Planck 
and BICEP2 data and show qualitatively that the power-law PPS is in fact ruled out at more than $3\sigma$. 
This is an important result, since it may require additional degrees of freedom to 
the current standard model of cosmology (assuming the observational results persist). 

At the same time most (slow-roll) inflationary scenarios also result to power-law form of the PPS and our findings show that all 
these models are now in fact in tension with observational data. In our analysis we study some well motivated phenomenological forms of the primordial spectrum such as a broken PPS model and a Tanh step model (we studied both models recently in  ~\cite{Hazra:2013nca}) as well as 
some potential theoretical models such as~\cite{Bousso:2013uia} to see if they can perform well fitting both Planck and BICEP2 data. 
Assuming these models and by deviating from some basic assumptions such as inflationary consistency relation we study how well we  can address the combined data and what are the affects on the constraints of the cosmological parameters. 
We show that step model and model with a break in scalar PPS are suitable candidates to fit both Planck and BICEP2 data and this  motivates some particular inflationary model  building. 
We perform full Monte Carlo sampling in our analysis to estimate the consistency of the the power-law form of PPS with Planck and BICEP2 data and for some other cases we limit ourselves to some particular model samples when we study inflationary scenarios. 
In a companion paper we focus on inflationary scenarios that may be able to fit all different data satisfactorily.

This paper is organized as follows. We discuss first the assumed phenomenological models and 
then we test the consistency of the power-law scalar PPS with Planck and combination of Planck and BICEP2 data. 
We then briefly discuss about potential inflationary scenarios that can fit both Planck and BICEP2 data. 
We end the paper by results and conclusions.  
\section{Formalism}~\label{sec:formalism}
In this work we have used the two scalar PPS that we had used in Ref~\cite{Hazra:2013nca} along with a tensor PPS. For the first model we have used the model-C with two bins 
which allows two spectral tilts in two bins and rename it as broken PPS + r for convenience. We define a scale $k_{\rm b}$ which indicates the break in the scalar PPS. Before
the break the scalar PPS is assumed to have a tilt $n_{\rm S1}$ and beyond that the spectral tilt is denoted by $n_{\rm S2}$. The amplitude of the scalar PPS 
is defined at the break and is denoted by $A_{\rm S}$. This shape of the model is similar to the exact solution for the scalar PPS 
found in~\cite{JSS08} (see also~\cite{JSSS09}).

The second phenomenological model considers a Tanh step at scale $k_{\rm b}$ in the scalar PPS and given by 
Eq.~\ref{eq:tanh} (also discussed in~\cite{Hazra:2013nca}).  
\beq
P_{\rm S}^{\rm Tanh}(k) = P_{\rm S}^{\rm Plaw}(k)\times\l[1+\alpha\tanh\l[\frac{k-k_{\rm b}}{\Delta}\r]\r]~\label{eq:tanh}
\eeq
\noindent
Here, $\alpha$ and $\Delta$ are the height and width of the step respectively. Note that this step is applied to the conventional power law primordial $P_{\rm S}^{\rm Plaw}(k)$
spectrum with amplitude $A_{\rm S}$ and tilt $n_{\rm S}$. In turn, this model is a simplified and smoothed version of the exact scalar PPS derived in~\cite{S92}. 
A similar PPS was also studied in~\cite{CPKL03}.

We should mention that for slowly rolling inflaton, consistency relation fixes the tensor spectral index $n_{\rm T}$  to have a small red tilt through $r=-8 n_{\rm T}$.
It can be argued that a strong blue tilt in the tensor PPS may address the inconsistencies between Planck and BICEP2 data since a blue tilt will allow 
negligible tensor contribution at low-$\ell$ and address the Planck data and around the bump in the BICEP2 BB-spectra the tensor contribution will be adequate to address the 
BICEP2 data as well. Now, for a fast roll during inflation, $n_{\rm T}$ can deviate from the consistency relation locally, but a global violation of consistency relation 
in the canonical framework should be dealt with a change in the scalar PPS that can restore it. For both the scalar PPS described above, we have performed our analysis 
by fixing tensor PPS tilt following consistency relation and by allowing it to vary.

For a model from inflationary theory that can represent the broken and the step like scalar PPS we use the model which 
appeared in the paper~\cite{Bousso:2013uia}.  
It was motivated by decay from a false vacuum to later inflation.
The potential essentially constitutes of a rapidly varying part, $V_R(\phi)$ and a slowly varying part 
$V_S(\phi)$ and is given by 
\beq
V(\phi)=V_S(\phi)+\gamma\times V_R(\phi),~\label{eq:potential}
\eeq
where, $\gamma$ denotes the strength of the rapid part of the potential. 
The rapid and the slow parts of the potential are given by,

\begin{eqnarray}
V_S(\phi)&=&V_i~(1-\sqrt{2\beta}\phi)\nn\\
V_R(\phi)&=&V_i~\Theta(\phi_c-\phi)~\frac{(\phi_c-\phi)^\zeta}{\zeta},~\label{eq:potential_rs} 
\end{eqnarray}
where, $\Theta(\phi_c-\phi)$ is the Heaviside theta function that turns off the rapid varying potential after a field value $\phi_c$. 
The initial stage of inflation is dominated by the rapid part of the potential following a power law (with power $\zeta>1$). Due to the fast-roll to slow-roll 
transition, the modes leaving the Hubble scale shall imprint the effect in the scalar power spectra. Interestingly, this potential can generate 
the phenomenological broken and the Tanh step scalar PPS in two different limits of $\zeta$. We search for the best fit in two different directions
of $\zeta$ to find the similar to the scalar PPS described above. We should mention that in this model the tensor-to-scalar ratio obtained is 
approximately $0.07-0.1$, which is lower than needed to address the BICEP2 peak of BB angular power spectra. However this model is interesting, 
since it is able to provide a balance between Planck low-$\ell$ suppression and BICEP2 B-mode data. For the evaluation of background and perturbation equations 
for the potential we have used the publicly available code {\tt BINGO}~\cite{Hazra:2012yn}. 

The potential that we are using from~\cite{Bousso:2013uia} is very similar to~\cite{S92} in the limit $\zeta=1$. The paper~\cite{Bousso:2013uia}
discussed inflation with first order phase transition at the GUT scale and the formation of Coleman - de Luccia bubbles, following the 
previous papers on this topic~\cite{Linde:1998iw,Linde:1999wv}. This is also similar to what was originally proposed in~\cite{Sato:1980}, but followed 
by $N\sim 60$ e-folds of more standard slow-roll inflation. Moreover this type of transition can also be achieved by coupling a massive field to 
inflaton~\cite{S92,JSS08,JSSS09,S98}. 

Through an intermediate fast roll, it has been demonstrated that features in the scalar PPS can be generated and can address the data 
better~\cite{features,Hazra:2010ve} than the power law form if scalar PPS. In particular, Ref.~\cite{Hazra:2010ve} had specifically calculated 
the tensor PPS along with the scalar PPS for canonical and non-canonical scalar field models with localized features without any approximation and 
provided a complete analysis with CMB datasets for different tensor contributions and highlighting the differences.

  We should mention that throughout our analysis we have defined $r = r_{0.05}$ at the pivot scale of $0.05 {\rm Mpc^{-1}}$.
  We have used publicly available software {\tt CAMB}~\cite{cambsite,Lewis:1999bs} to generate the angular power spectrum and 
  {\tt COSMOMC}~\cite{cosmomcsite,Lewis:2002ah} with Planck likelihood code to perform the Markov Chain Monte Carlo (MCMC) analysis. We have 
  used Planck angular power spectrum for temperature anisotropy, low-$\ell$ (2-23) polarization data from WMAP and BICEP2 data (E and B mode). We have used 
  BICEP2 bandpowers for 9 bins. For the analysis with the 
  BICEP2 data we have used the newHL branch of {\tt COSMOMC}. Along with the scalar and tensor PPS and background cosmological parameters we do also 
  marginalize over the 14 nuisance parameters corresponding to different foreground and calibration effects in different frequency channels. Marginalization over
  all the underlying likelihood nuisance parameters make our analysis robust. For finding the best fit parameters and the 
  likelihood we have used Powell's BOBYQA method of iterative minimization~\cite{powell}. In all our analyses we have assumed spatially flat FLRW universe. 
  We have defined ${\rm M_{PL}}^2=1/(8\pi G)$ and used $\hslash=c=1$ throughout the paper.

\section{Results and discussions}\label{sec:results}

\begin{figure*}[!htb]
\vskip -60pt
\begin{center} 
\resizebox{210pt}{160pt}{\includegraphics{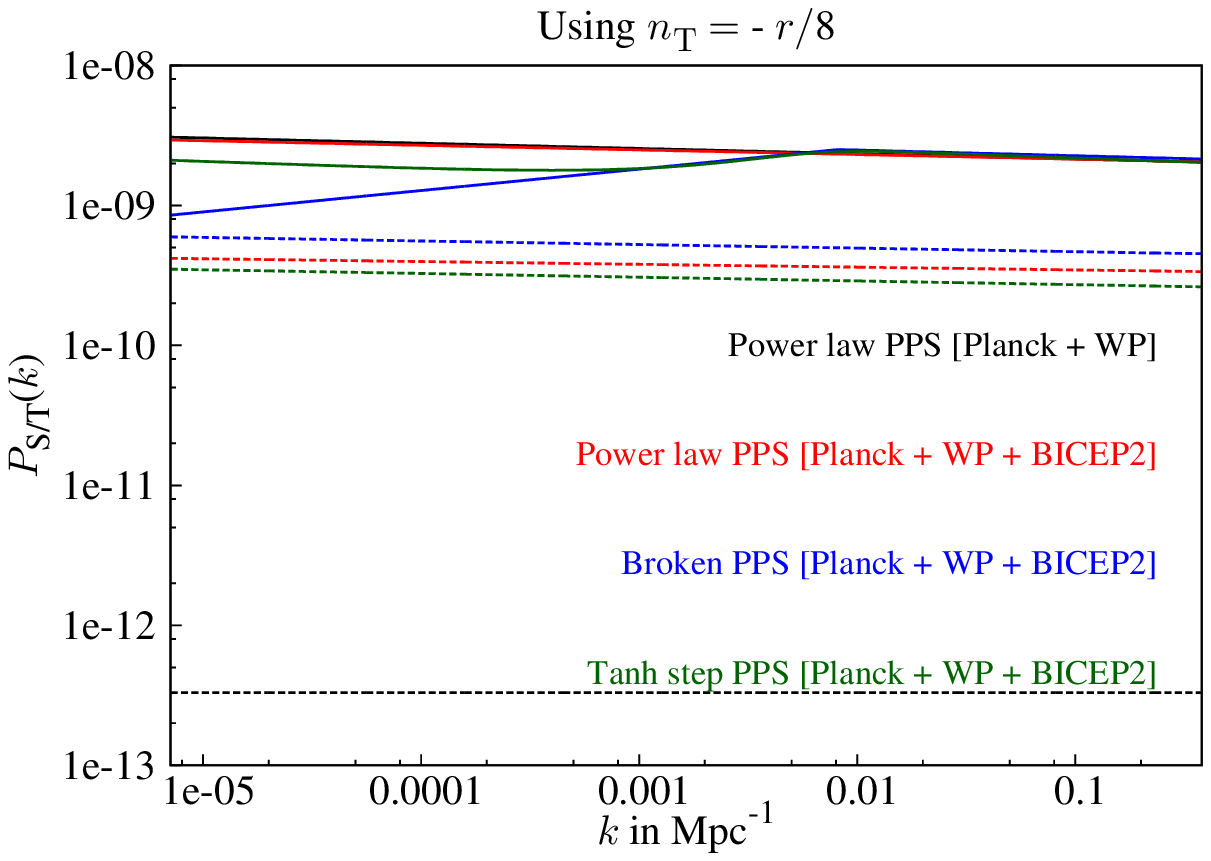}} 
\resizebox{210pt}{160pt}{\includegraphics{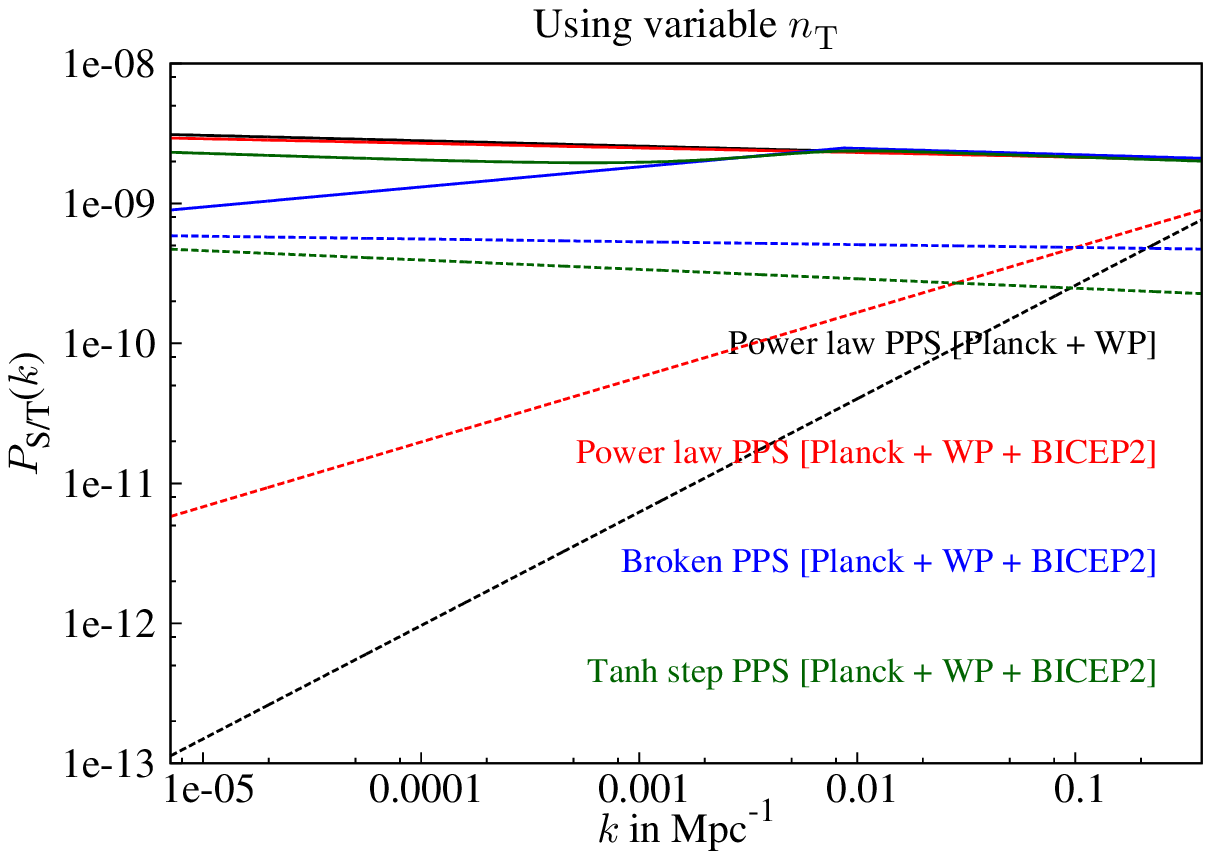}} 
\resizebox{210pt}{160pt}{\includegraphics{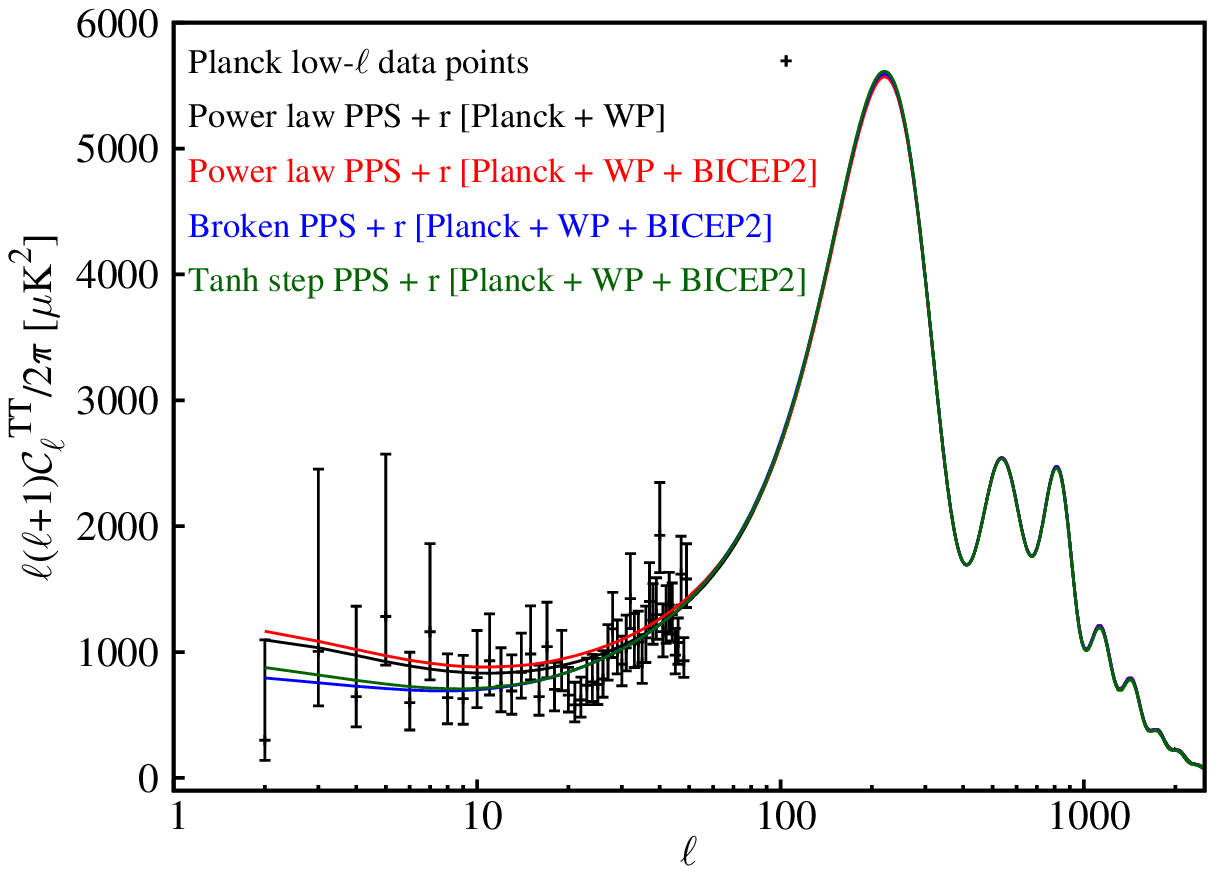}} 
\resizebox{210pt}{160pt}{\includegraphics{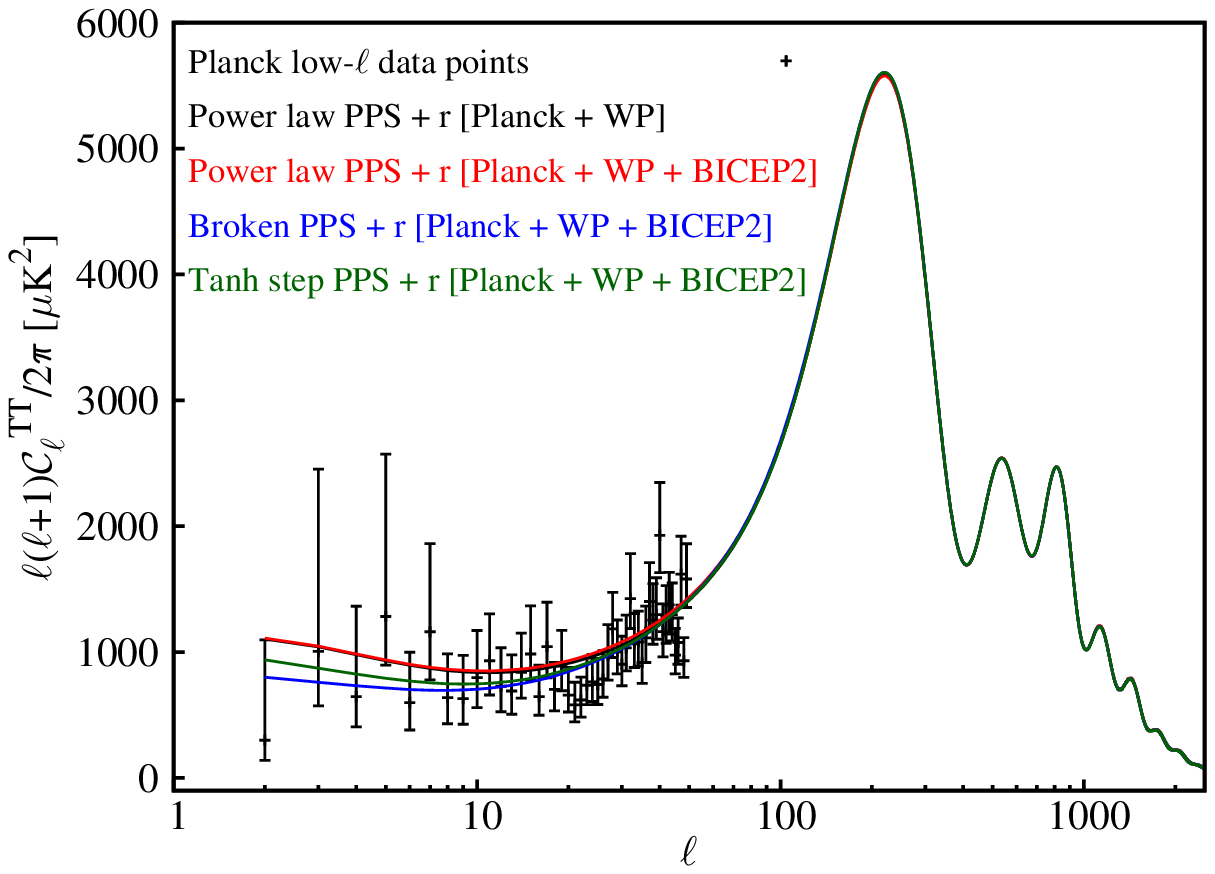}} 
\resizebox{210pt}{160pt}{\includegraphics{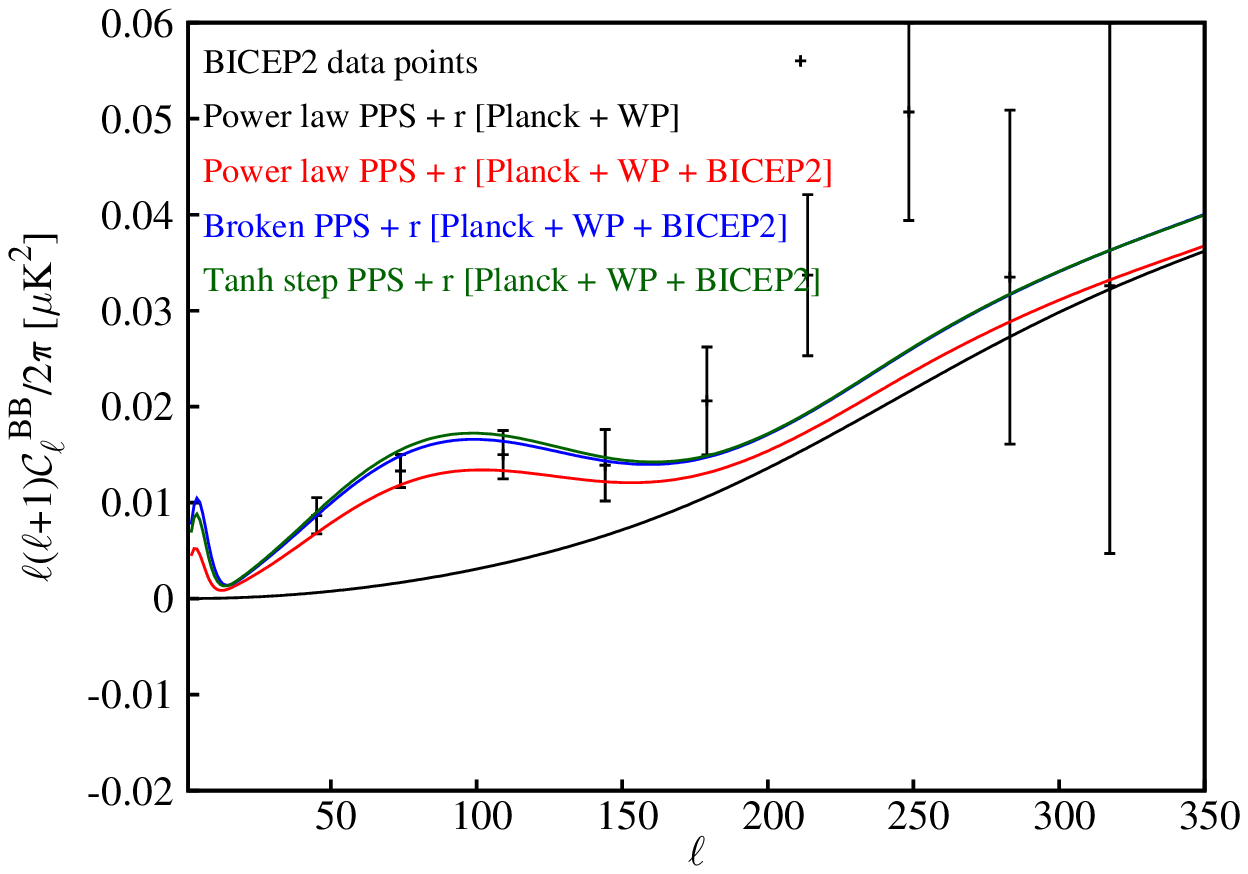}} 
\resizebox{210pt}{160pt}{\includegraphics{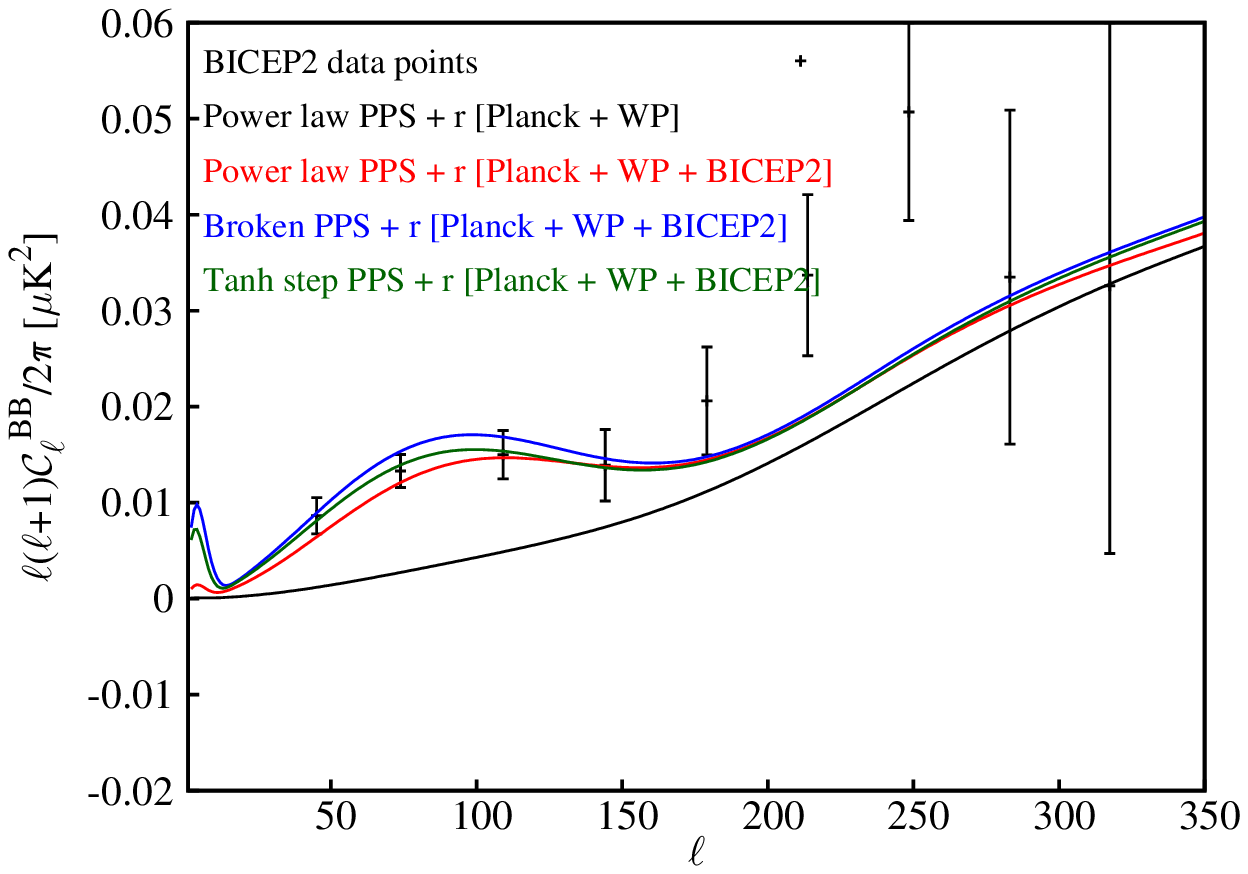}} 

\end{center}
\caption{\footnotesize\label{fig:bestfiticon} Left : Results obtained when we assume inflationary consistency relation.
Best fit primordial power spectra obtained in different 
analyses are plotted at the top panel. Scalar PPS is plotted in solid and tensor PPS are plotted in dashed lines. 
Note that when we add BICEP2 data, for power law form of PPS tensor PPS goes to higher amplitude (compare black and red dashed line). 
However, a break (blue solid curve) or a Tanh step (green solid curve) in the scalar PPS indicates that it can 
address the data better than power law form with even higher 
tensor PPS. Best fit angular power spectra of temperature anisotropy obtained in different 
 analyses are plotted in the middle. Note that when we add BICEP2 data, the power law form of PPS is not able to fit the 
 data well (see the black curve) and provides a  worse likelihood in low-$\ell$ compared to the red curve. 
 However, a break (blue curve) or step (green curve) in the PPS fits the low-$\ell$ Planck data better than the 
 other two curves. Best fit angular
 power spectra of B-mode polarization obtained in different 
 analyses and the BICEP2 data are plotted at the bottom. Note that the black curve for best fit power law from Planck + WP
 is unable to address the bump in the BB angular power spectra. The broken and Tanh scalar PPS
 fits the Planck large scale data significantly better than power law and allows $r$ to have a large value which at the same
 time fits BICEP2 data better. Right : the same primordial and angular power spectra are plotted for the case where tensor spectral 
 index is allowed to vary. A blue tensor spectral tilt (red dashed curve in the top right panel) helps to reduce 
 the increase in $\cl^{\rm TT}$ at largest scales and provide the adequate tensor contribution to fit the BB bump around $\ell\sim100$. 
 The break and the step model does not require $n_{\rm T}$ to have a large blue tilt and restores inflationary consistency relation.}
 \end{figure*}

We begin by presenting the best fit results obtained for power law and the two modified scalar PPS in Fig.~\ref{fig:bestfiticon}. 
The left panels consist the result when we fix the tensor spectral index following inflationary consistency relation and the right 
panel represent the results when we treat $n_{\rm T}$ as free parameter. 

The top panels consist of best fit primordial scalar and 
tensor power spectra in different cases. 
Note that when we do not include BICEP2 data, the tensor power spectra in both the cases 
are significantly smaller than the cases where we include BICEP2 data\footnote{The black dashed line for tensor PPS in the right 
hand side indicates a high blue tilt for Planck + WP in power law case. Note that this is a random choice in the $n_{\rm T}-r$ 
parameter space since at large scales we can decrease tensor contribution to $\cl^{\rm TT}$ either by decreasing $r$ or by 
adding a blue tilt in tensor PPS}. 
The BICEP2 data increases the power of the tensor PPS in all the cases. The middle panel contains the $\cl^{\rm TT}$ for the same scalar
and tensor PPS plotted in the upper panel and the Planck low-$\ell$ (2-49) $\cl^{\rm TT}$ data. It is clear from the figure that when we 
add BICEP2 data the resulting $\cl^{\rm TT}$ from the best fit scalar and tensor PPS fits Planck low-$\ell$ $\cl^{\rm TT}$ data worse than 
the best fit power law PPS from Planck + WP only, since the large tensor component (as demanded by BICEP2 data) increases the $\cl^{\rm TT}$ at large scales. 
When we allow $n_{\rm T}$ to vary, the data combination prefers a large blue tilt in the tensor PPS for the power law model (red dashed 
line in the top tight panel) which allows 
negligible tensor contribution to TT spectrum (red and black curves in the middle right plots match) 
and adequate tensors around the bump in BB. The bottom panels contain the BB data from BICEP2 
and the best fit $\cl^{\rm BB}$ for different models. Note that best fit models obtained fitting Planck + WP does not address BB data at all. 
In all the cases we note that the scalar PPS models with a break and a Tanh step fits both the data better than power law model. Both the models 
allow a significant drop in the scalar PPS which fits $\cl^{\rm TT}$ data better than the power law and also allow tensors to be large to 
fit the BB data at the same time. Interestingly, note that when we use a break or a step in the scalar PPS, the drop in power at large scales 
relaxes the violation of consistency relation to a considerable extent. For example the 
 \clearpage
\noindent blue and the green dashed lines in the top right plot indicates a nearly scale invariant tensor PPS.

The best fit values of the cosmological parameters and the likelihoods corresponding to the models are tabulated in Table~\ref{tab:ic-on},~\ref{tab:ic-off}\footnote{The 
* marks in the scalar spectral amplitude for the broken PPS model indicate that the amplitude is defined at the break of the PPS.} 
and in Table~\ref{tab:Tanh}. Table~\ref{tab:ic-on} contains the best fit values for power law and the broken scalar PPS when we fix
$n_{\rm T}=-r/8$. Table~\ref{tab:ic-off} contains results for the same models but allowing tensor spectral index to vary. Table~\ref{tab:Tanh}
contains the results for the Tanh step model for fixed and variable $n_{\rm T}$. $-2\Delta\ln{\cal L}$ in each table indicates the difference 
in log likelihood (${\cal L}$) between different modified models and the power law model for the corresponding combinations of datasets and for the same 
assumption of the tensor spectral index. The breakdown of $\ln{\cal L}$ in different datasets for the best fits are also provided. 
Note that when BICEP2 data is added to Planck + WP combination the power law model fits the Planck low-$\ell$ data worse by $-2\Delta\ln{\cal L}\sim6$ 
(comparing {\tt commander} likelihoods). However, the broken and the Tanh step scalar PPS models fits the Planck low-$\ell$ data better than power 
law and at the same time allow $r$ to be larger and fits BICEP2 likelihood better too. The break and the step models increases the overall fit to the 
complete datasets by ${\cal O}(12-13)$ compared to power law model. Table~\ref{tab:ic-off} and the Table~\ref{tab:Tanh} (the part of the table where 
$n_{\rm T}$ is varied) indicate that $n_{\rm T}$ does not need to have a blue tilt when we allow a suppression in large scale scalar PPS.  
Moreover Table~\ref{tab:ic-off} and Table~\ref{tab:Tanh} specifically show that when we are allowing the scalar 
power suppression through a broken or step PPS, allowing $n_{\rm T}$ to vary does not result to significant improvement in the log likelihood, 
which clearly states that the consistency relation does not need to be violated.
 
\renewcommand{\arraystretch}{1.1}

\begin{table*}[!htb]
\begin{center}
\vspace{4pt}
\begin{tabular}{|c | c | c || c | c |}
\hline\hline
 \multicolumn{5}{|c|}{{\bf Comparison of the broken scalar PPS with power law spectra}}\\

 \hline
 & \multicolumn{2}{|c||}{{\bf Planck + WP}}& \multicolumn{2}{|c|}{{\bf Planck + WP + BICEP2}}\\
\cline{2-5}

$n_{\rm T}=-r/8$& Power law PPS & Broken PPS & Power law PPS & Broken PPS \\

\hline

$\Omega_{\rm b}h^2$ & 0.02207 & 0.02192 & 0.02205&0.02201\\
\hline

$\Omega_{\rm CDM}h^2$ & 0.1197 &  0.1225&0.1189 &0.1204\\
\hline

$100\theta$ & 1.041 & 1.041 &1.041 &1.041\\
\hline

$\tau$ & 0.091 & 0.096 & 0.09 &0.117\\
\hline

$n_{\rm S1}$ & 0.9625 & 1.041 & 0.9672&1.15\\
\hline

$n_{\rm S2}$ & - & 0.9521 & -&0.9594\\
\hline

$r$ & $1.5\times10^{-4}$ & 0.03 & 0.16&0.204\\
\hline

$k_{\rm b}$ & - & 0.009 & -&0.008\\
\hline

${\rm{\ln}}(10^{10} A_{\rm S})$ & 3.093 & $3.19^\ast$ &3.088 &$3.22^\ast$\\

\hline\hline

$\Omega_{\rm m}$ & 0.314 & 0.33 & 0.31&0.32\\
\hline
$H_{0}$ & 67.3 & 66.1 & 67.6&67\\
\hline\hline
\multicolumn{5}{c}{$-2\ln{\cal L}$ [Best fit]}\\
\hline
{\tt commander} & -6.98 & -10.1 &-1.13 &-9.66\\
\hline
{\tt CAMspec} & 7795.13 & 7794 & 7797.29&7793.55\\
\hline
{\tt WP} & 2014.34 & 2014.56 & 2013.38&2014.66\\
\hline
{\tt BICEP2} & - & - &40.04 &38.4\\
\hline
Total & 9802.49 & 9798.46 &9849.58 &9836.95\\
\hline
$-2\Delta\ln{\cal L}$ & - & -4.01 & - &-12.63\\
\hline

\hline\hline
\end{tabular}
\end{center}
\caption{~\label{tab:ic-on} Table of cosmological parameters with fixing tensor spectral index to $n_{\rm T}=-r/8$, satisfying 
inflationary consistency relation. we compare the best fit parameter and likelihoods obtained from assuming power law primordial power 
spectrum and assuming a power spectrum broken at a particular scale $k_{\rm b}$ ${\rm Mpc^{-1}}$. The broken power spectrum is characterized by 
two spectral tilt namely, $n_{\rm S1}$ and $n_{\rm S2}$. Note that the broken power spectrum is providing a significant improvement in fit 
compared to power law model. Moreover note that the break in the spectrum helps to fit both Planck and BICEP2 data better. The improvement 
in {\tt commander} likelihood indicates that the break in the scalar power spectra allows us to fit large scale angular power spectrum from Planck 
and at the same time allowing a higher value of $r$ helps fitting the BICEP2 data better. 
Tensor power spectrum is assumed to be power law in all the cases.}
\end{table*}

\renewcommand{\arraystretch}{1.1}

\begin{table*}[!htb]
\begin{center}
\vspace{4pt}
\begin{tabular}{|c | c | c || c | c |}
\hline\hline
 \multicolumn{5}{|c|}{{\bf Comparison of the broken scalar PPS with power law spectra}}\\

 \hline
& \multicolumn{2}{|c||}{{\bf Planck + WP}}& \multicolumn{2}{|c|}{{\bf Planck + WP + BICEP2}}\\

\cline{2-5}

Variable $n_{\rm T}$ & Power law PPS & Broken PPS & Power law PPS & Broken PPS \\

\hline

$\Omega_{\rm b}h^2$ & 0.022& 0.0222&0.022 &0.022\\
\hline

$\Omega_{\rm CDM}h^2$ & 0.1197 & 0.1185 & 0.1193&0.1208\\
\hline

$100\theta$ &  1.041& 1.041& 1.041&1.041\\
\hline

$\tau$ &0.093  & 0.102 & 0.09&0.112\\
\hline

$n_{\rm S1}$ &  0.9615& 1.099 & 0.9633&1.114\\
\hline

$n_{\rm S2}$ &  -& 0.9644& -&0.9554\\

\hline
$r$ &  0.067& 0.085& 0.377&0.2137\\

\hline
$n_{\rm T}$ & 0.8 & -0.038& 0.463&-0.02\\

\hline

$k_{\rm b}$ & -& 0.0055& -&0.0087\\
\hline

${\rm{\ln}}(10^{10} A_{\rm S})$ &3.095 &$3.19^{\ast}$  & 3.088&$3.2^{\ast}$\\
\hline\hline

$\Omega_{\rm m}$ & 0.315 &  0.306& 0.312&0.32\\
\hline
$H_{0}$ &  67.3& 67.9& 67.4&66.8\\
\hline\hline
\multicolumn{5}{c}{$-2\ln{\cal L}$ [Best fit]} \\
\hline
{\tt commander} & -6.63 &-10  & -4.51&-9.69\\
\hline
{\tt CAMspec} & 7795.16 & 7795.68& 7796.4&7794\\
\hline
{\tt WP} & 2014.43 & 2013.95& 2013.95&2014.07\\
\hline
{\tt BICEP2} & - &  -& 38.21&38.47\\
\hline
Total &  9802.96& 9799.63& 9844.05&9836.85\\
\hline
$-2\Delta\ln{\cal L}$ & - & -3.33 & - &-7.2\\
\hline

\hline\hline
\end{tabular}
\end{center}
\caption{~\label{tab:ic-off}Table of cosmological parameters with allowing tensor spectral index $n_{\rm T}$ to vary. Note that the break in 
the scalar power spectra allows $n_{\rm T}$ to have a very small value and restores the inflationary consistency relation. It is interesting to 
notice that a blue tensor spectral index with power law scalar PPS can help to fit both Planck low-$\ell$ and BICEP2 data better than the case 
where we assume inflationary consistency relation. In fact a large blue tilt in the primordial tensor spectrum suppresses the effect of high $r$
on $\cl^{\rm TT}$ at largest scales probed by Planck and around $\ell\sim100$ fits the bump in the $\cl^{\rm BB}$ data. However, compared to the 
broken scalar PPS, the effect of $n_{\rm T}$ is rather limited as can be seen by comparing the {\tt commander} likelihoods obtained in different
cases.}
\end{table*}

\renewcommand{\arraystretch}{1.1}

\begin{table*}[!htb]
\begin{center}
\vspace{4pt}
\begin{tabular}{|c | c | c || c | c |}
\hline\hline
 \multicolumn{5}{|c|}{{\bf Comparison of the Tanh step scalar PPS with power law spectra}}\\

 \hline
& \multicolumn{2}{|c||}{{\bf Planck + WP}}& \multicolumn{2}{|c|}{{\bf Planck + WP + BICEP2}}\\

\cline{2-5}

Tanh Model& $n_{\rm T}=-r/8$ & Variable $n_{\rm T}$ & $n_{\rm T}=-r/8$ & Variable $n_{\rm T}$ \\

\hline

$\Omega_{\rm b}h^2$ & 0.0219& 0.0218&0.0218 &0.0218\\
\hline

$\Omega_{\rm CDM}h^2$ & 0.1208 & 0.1222 & 0.1222&0.1226\\
\hline

$100\theta$ &  1.041& 1.041& 1.041&1.041\\
\hline

$\tau$ &0.105  & 0.087 & 0.103&0.092\\
\hline

$\alpha$ &  0.121& 0.115 & 0.746&0.5\\
\hline

$\ln\Delta$ & -9.41& -9.4 &-5.3&-5.2\\
\hline

$n_{\rm S}$ &  0.9552& 0.9478& 0.9454&0.9491\\

\hline
$r$ &  0.03& 0.0002& 0.213&0.1756\\
 
 \hline
 $n_{\rm T}$ & - & -0.16& -&-0.067\\

\hline

$k_{\rm b}$ & 0.0028& 0.0028& $1.3\times10^{-5}$&$1.1\times10^{-5}$\\
\hline

${\rm{\ln}}(10^{10} A_{\rm S})$ &3.08 &2.56  & 2.937&2.69\\
\hline\hline

$\Omega_{\rm m}$ & 0.32 &  0.33& 0.33&0.33\\
\hline
$H_{0}$ &  66.8& 66.0& 66.2&66.0\\
\hline\hline
\multicolumn{5}{c}{$-2\ln{\cal L}$ [Best fit]} \\
\hline
{\tt commander} & -12.11 &-12.06  & -10.1&-9.93\\
\hline
{\tt CAMspec} & 7794.44 & 7795.07& 7795.42&7794.6\\
\hline
{\tt WP} & 2015.22 & 2014.91& 2013.77&2013.54\\
\hline
{\tt BICEP2} & - &  -& 38.79&39.07\\
\hline
Total &  9797.55& 9797.92&9837.59&9837.28\\
\hline
$-2\Delta\ln{\cal L}$ & -4.94 & -5.04 & -12 &-6.77\\
\hline

\hline\hline
\end{tabular}
\end{center}
\caption{~\label{tab:Tanh}The best fit cosmological parameters for scalar PPS with Tanh step modification. Note that the parameter $\alpha$ which denotes the 
strength of the step is larger when we allow BICEP2 data that indicates a strong suppression in power needed in order to address Planck and BICEP2 data in the same framework.
The best improvement is found when we compare the results of power law scalar PPS and Tanh model while satisfying inflationary consistency relation. Allowing $n_{\rm T}$ 
to vary leads to an improvement in fit in the case of power law scalar PPS and thereby decreases the scope of further improvement by a step modification.}
\end{table*}

\begin{figure*}[!htb]
\begin{center} 
\hskip -10pt\resizebox{140pt}{120pt}{\includegraphics{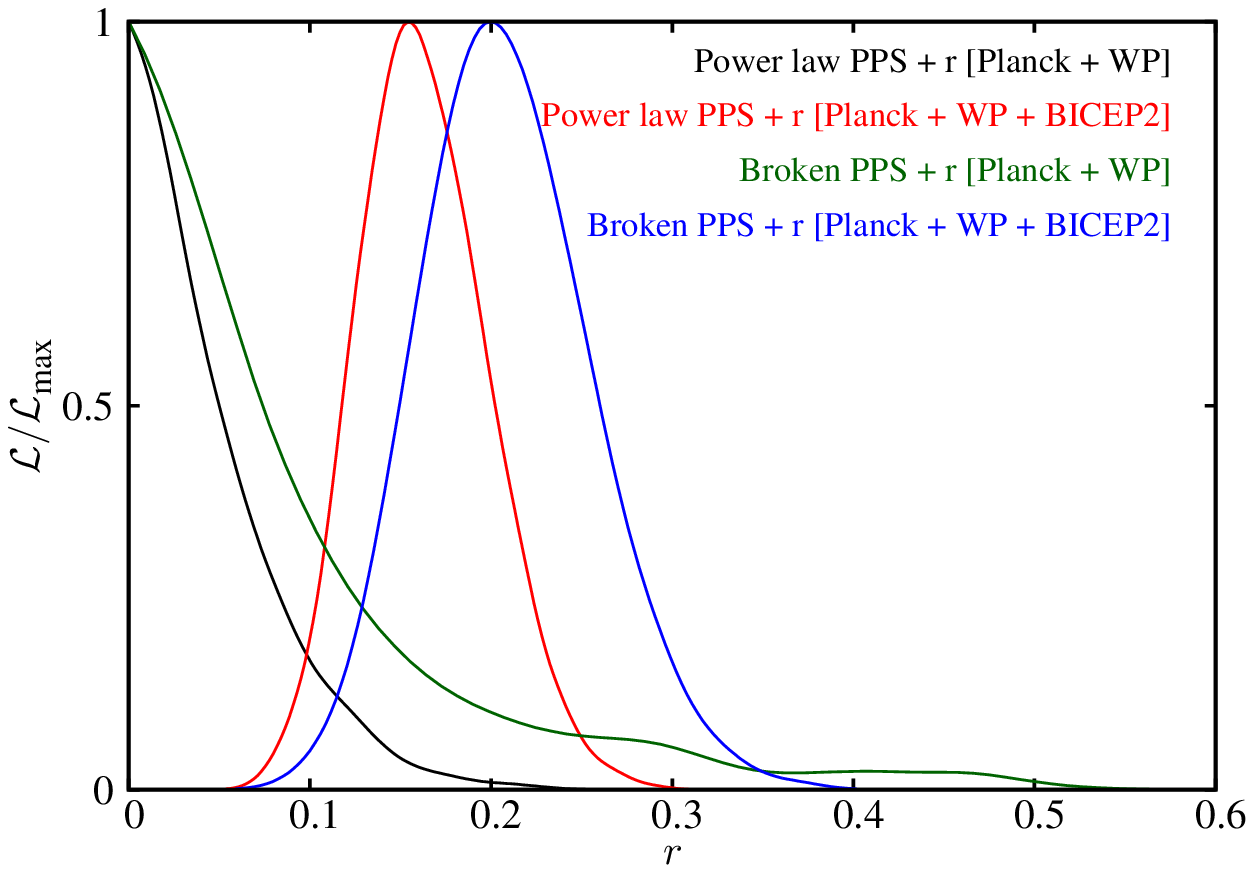}} 
\resizebox{140pt}{120pt}{\includegraphics{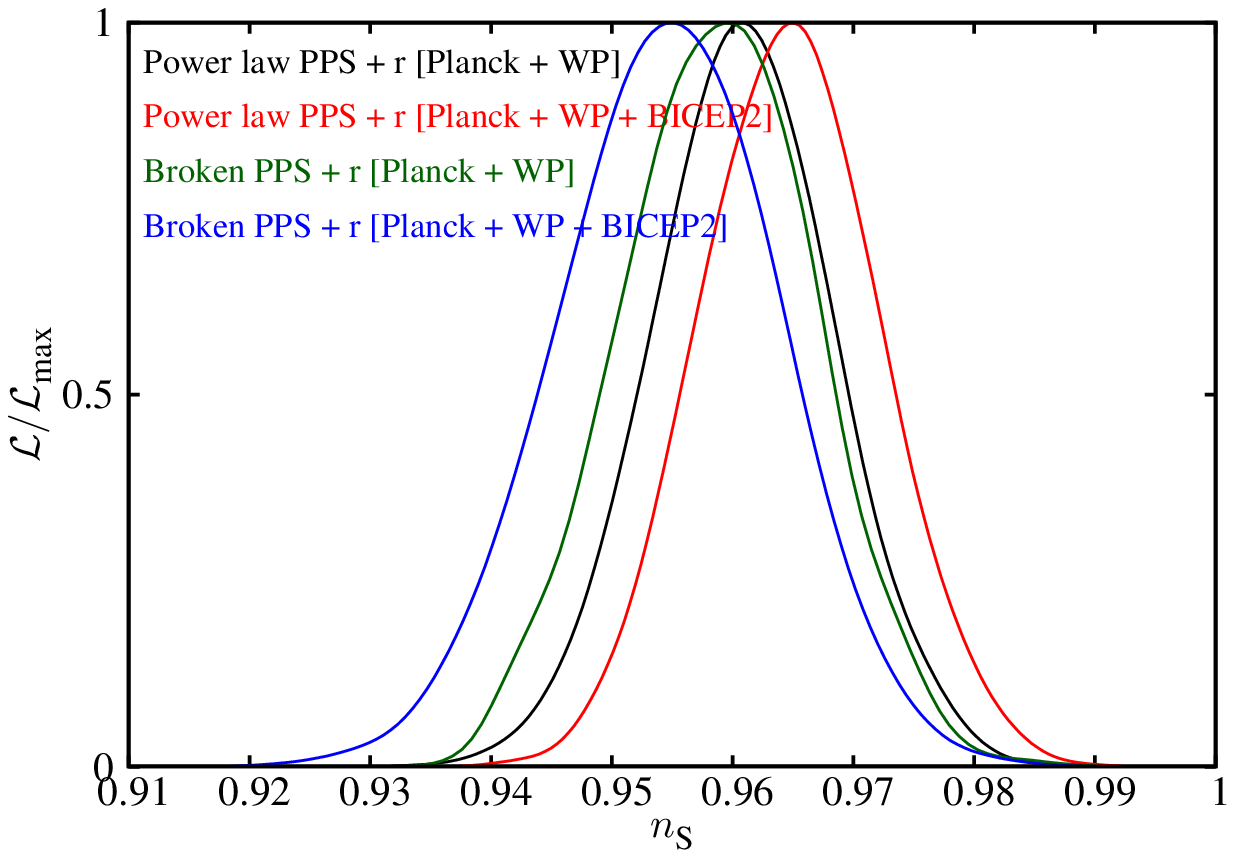}} 
\hskip -10pt \resizebox{160pt}{140pt}{\includegraphics{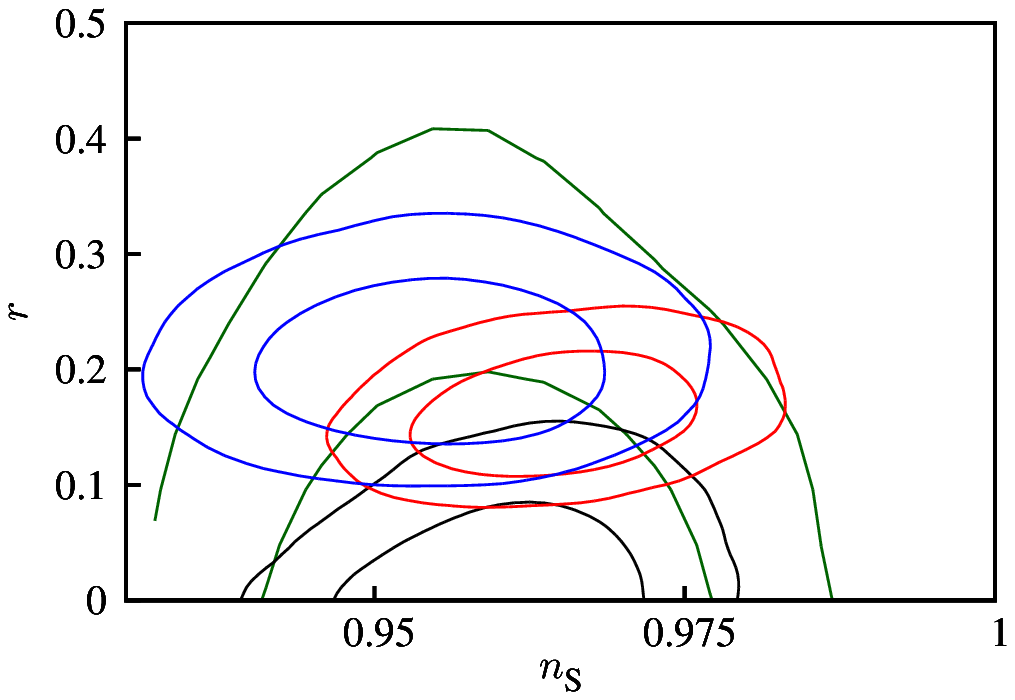}} 

\end{center}
\caption{\footnotesize\label{fig:icon1d2d} One dimensional marginalized contours of $r$ (left) and $n_{\rm S}$ (right). Note that for the broken PPS we have plotted 
the $n_{\rm S2}$, the tilt of the scalar PPS after the break. Here we have fixed the tensor spectral tilt using the inflationary consistency relation. Note that 
the broken power spectrum can address both the data from Planck and BICEP2 and can solve the inconsistencies within.}
\end{figure*}

Having described the best fit results for different models we now present the important contours of our analyses for the broken scalar PPS model. 
In Fig.~\ref{fig:icon1d2d} we plot the 1d and 2d marginalized likelihoods for the scalar spectral index ($n_{\rm S}$) and the tensor-to-scalar
ratio $r$. When we refer to the broken PPS in this plot the corresponding $n_{\rm S}$ denotes the spectral index of the scalar PPS after the break,
{\it i.e.} $n_{\rm S2}$. Note that the black and the red contours indicate an inconsistency between Planck + WP and Planck + WP + BICEP2 datasets.
Allowing a break we can resolve the inconsistency, as has been plotted in green curve. The blue contours shifts to higher $r$ direction compared to 
the red contours which supports the fact that improvement in fit is obtained through even higher $r$ which the broken PPS can allow by fitting
the Planck data substantially better.

\begin{figure*}[!htb]
\begin{center} 
\resizebox{210pt}{160pt}{\includegraphics{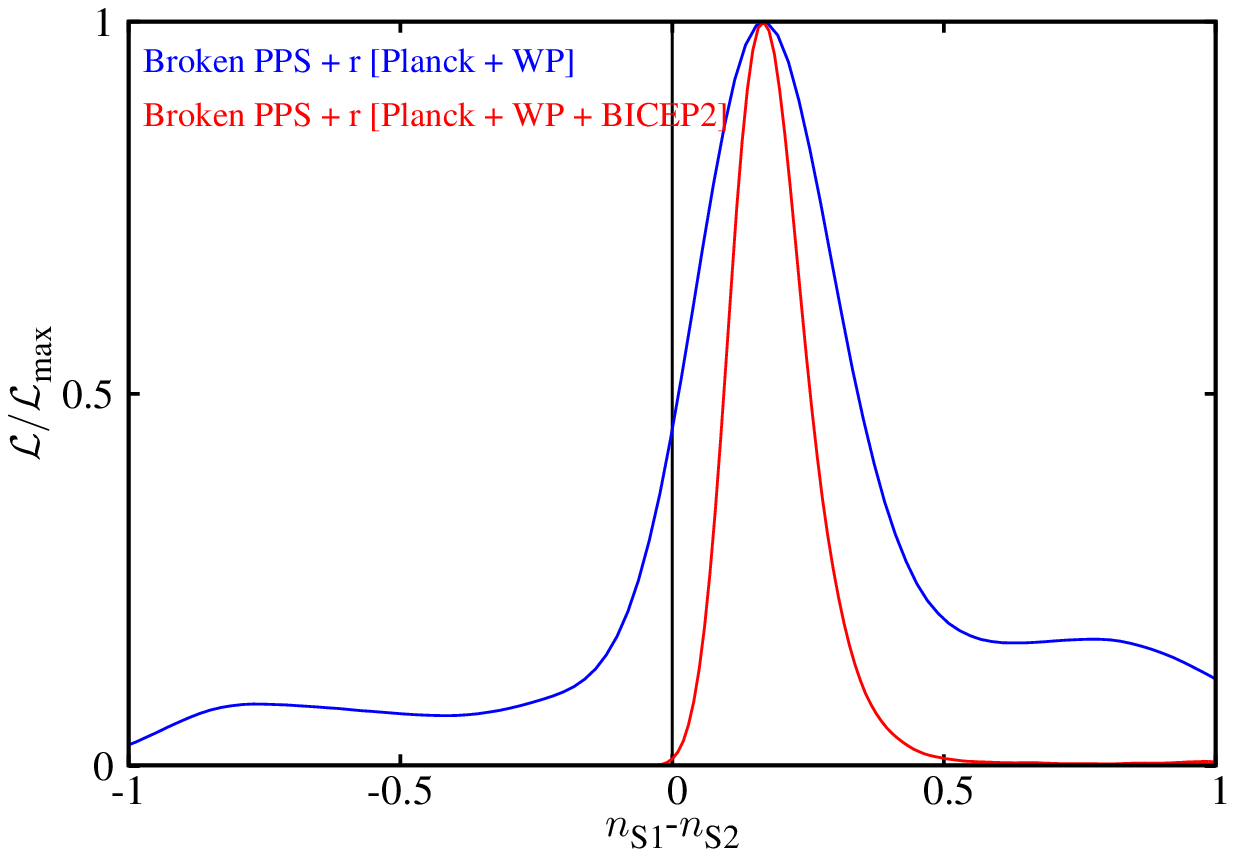}} 


\end{center}
\caption{\footnotesize\label{fig:ns1-ns2} The 1D marginalized likelihood obtained for the difference between the two spectral tilts ($n_{\rm S1}-n_{\rm S2}$) in the 
case of broken scalar PPS are plotted. The results shown from the analysis with Planck + WP and with Planck + WP + BICEP2 are plotted in blue and red respectively. The vertical
line at 0 in black represents the power law scalar PPS. Note that when we do not include BICEP2 data the power law is allowed within 1$\sigma$ C.L., but the addition
of BICEP2 data rules out the power law scalar PPS with high confidence (at more than 3$\sigma$ CL).}
\end{figure*}

Since a suppression in the scalar PPS is providing a substantial improvement in likelihood it is interesting to see the stand of power law in the light 
of Planck + WP and BICEP2 data combinations. In Fig.~\ref{fig:ns1-ns2} we plot the difference between the scalar spectral index before and after
the break ($n_{\rm S1}-n_{\rm S2}$) with (red) and without (blue) including BICEP2 data in the Planck + WP combination. Note that Planck + WP 
data though favors a blue tilt at large scales, it is not significant to rule out the power law scenario, corresponding to $n_{\rm S1}-n_{\rm S2}=0$.
However, when we add BICEP2 data, we find that power law is ruled out at more than 3$\sigma$ CL. 
This is a significant result since the cosmic variance
puts a limitation on the detection of the large scale features, 
but the BICEP2 data indirectly confirms the exclusion of power law model. 

\begin{figure*}[!htb]
\begin{center} 
\hskip -10pt\resizebox{140pt}{120pt}{\includegraphics{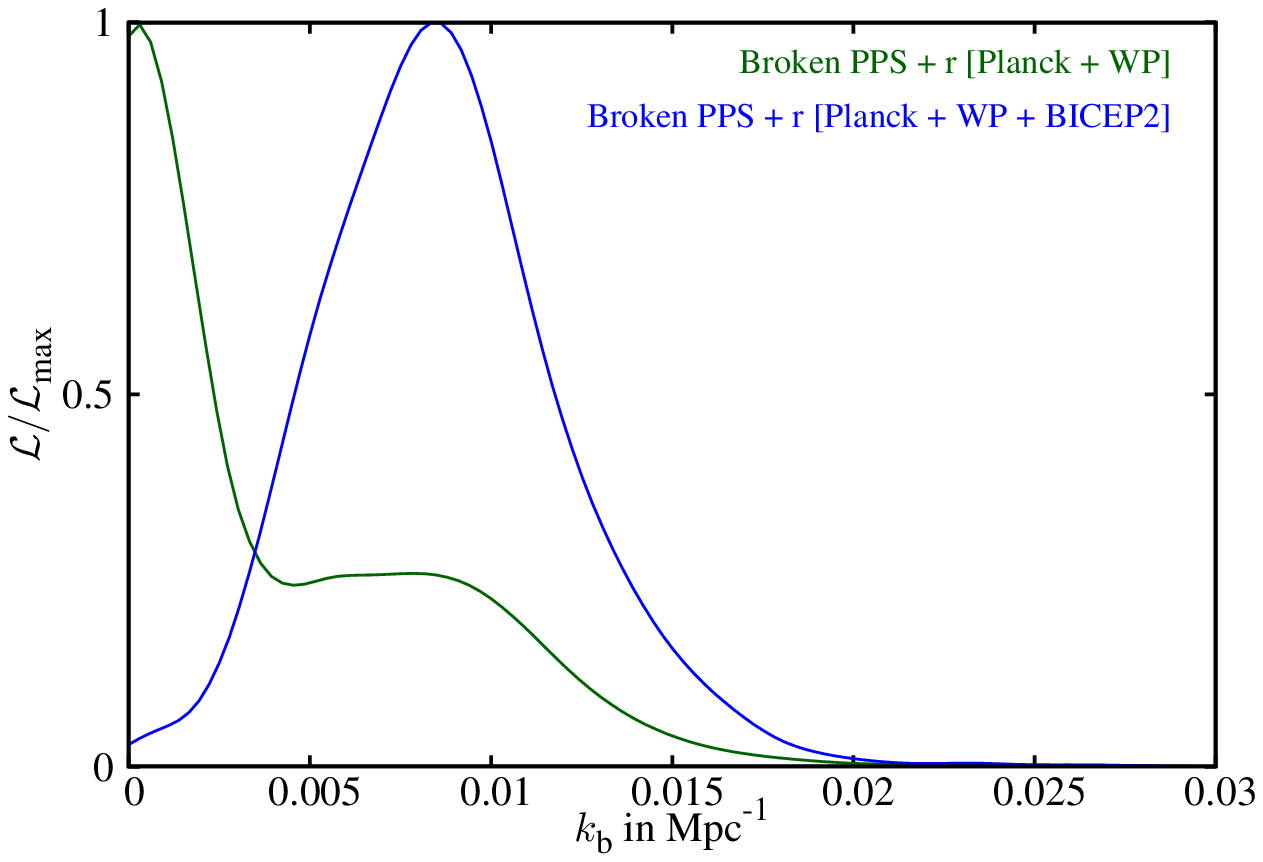}} 
\resizebox{140pt}{120pt}{\includegraphics{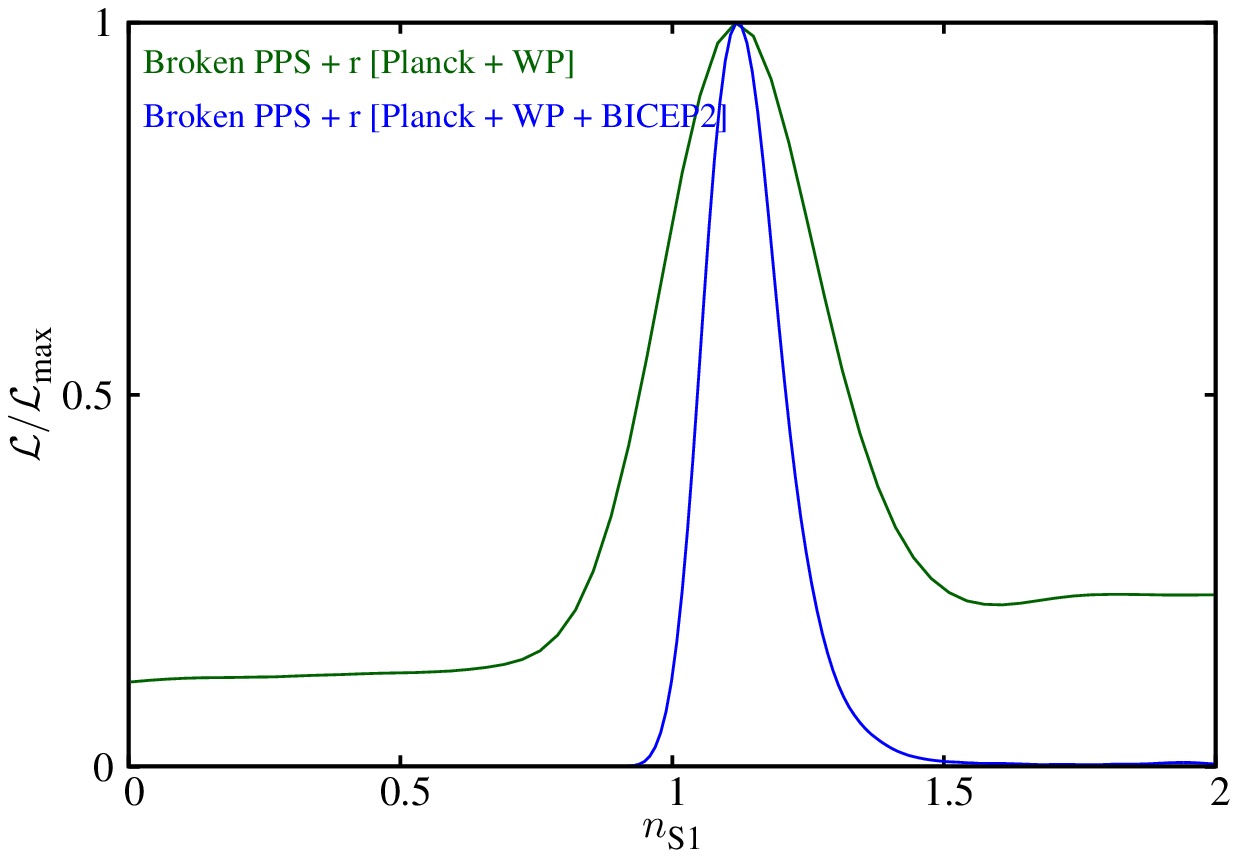}} 
\hskip -10pt \resizebox{160pt}{140pt}{\includegraphics{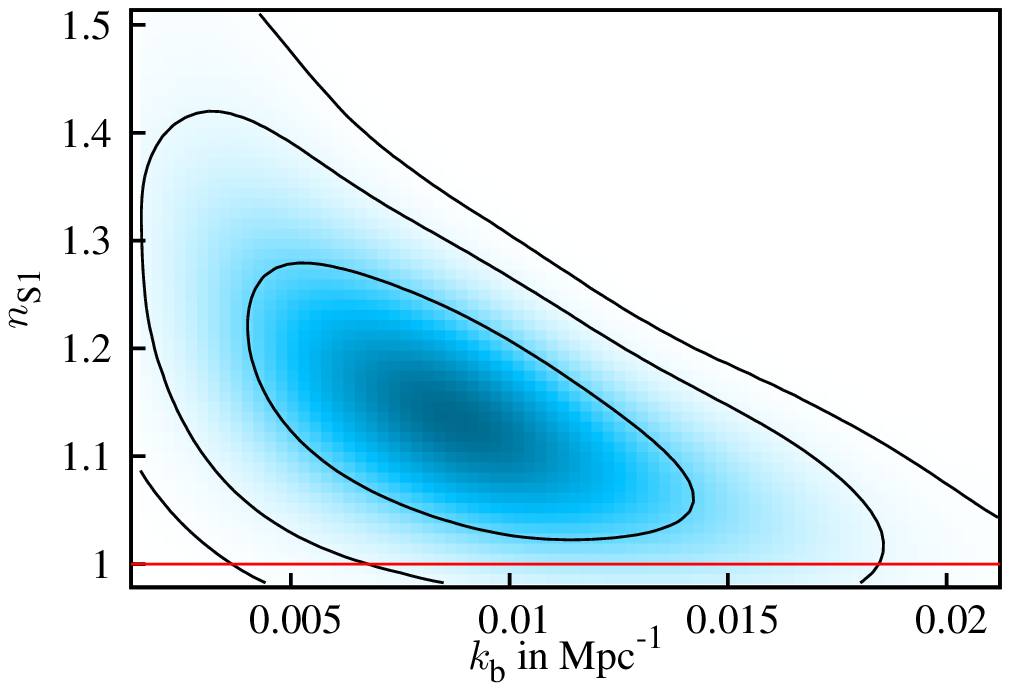}} 

\end{center}
\caption{\footnotesize\label{fig:kbns1} Left: One dimensional marginalized likelihoods for the position of the break ($k_{\rm b}$) in the scalar PPS. Middle: One dimensional 
marginalized likelihoods for the tilt of the scalar PPS before the break. Note that when we do not add the BICEP2 data (green curves), the  $k_{\rm b}=0$ (no break)
and $n_{\rm S1}$ with a red tilt is consistent with the data. However Planck + WP + BICEP2 combination forces $n_{\rm S1}$ to have a blue tilt at around a 
non-zero $k_{\rm b}$ to be consistent with Planck low-$\ell$ and BICEP2 data at the same time.
Right: Two dimensional marginalized probabilities of the break position in scalar PPS $k_{\rm b}$ and $n_{\rm S1}$. 
Note that in the first case a red tilt in scalar PPS at large 
scales (corresponding to $n_{\rm S1}<1$) is disfavored at more than 3$\sigma$ till 0.004 ${\rm Mpc^{-1}}$ and at more than 2$\sigma$ till 0.007${\rm Mpc^{-1}}$. This 
fact implies that to fit both the Planck and BICEP2 data reasonably we must have a drop in power at large scales. Comparing this plot with a 
similar plot of Ref.~\cite{Hazra:2013nca} indicates that with Planck + WP supports large scale drop in power at large scale but it was not evident
while this drop is now significantly favored with the addition of B-mode polarization data from BICEP2. Plots above represent the results where tensor 
spectral index fixed to $-r/8$.}
\end{figure*}

Since in all our analyses, we vary the position of the break $k_{\rm b}$, it helps us to hunt down the cosmological scales where we must need a break in the scalar PPS. In Fig.~\ref{fig:kbns1} 
we plot the 1d marginalized likelihoods of $k_{\rm b}$ (left) and $n_{\rm S1}$ (middle) for Planck + WP (green) and Planck + WP + BICEP2 (blue) dataset 
combinations. When we do not add BICEP2 we find that the $k_{\rm b}$  is consistent with 0 implying no break and $n_{\rm S1}$ consistent to 
a red tilt, confirming power law consistency. When we add BICEP2 data, the scenario changes and $k_{\rm b}$ is constrained within a particular 
window of cosmological scales that indicates the necessity of a break and $n_{\rm S1}$ rejects a value lower than 1 (red tilt) 
with high confidence. The scale dependence of this rejection is plotted in the 2d marginalized contours of $k_{\rm b}$-$n_{\rm S1}$ (right panel). Note that 
a red tilt is disfavored at more than 3$\sigma$ till $\sim0.004 {\rm Mpc^{-1}}$ and at more than 2$\sigma$ till $\sim0.007 {\rm Mpc^{-1}}$. 

\begin{figure*}[!htb]
\begin{center} 
\resizebox{210pt}{160pt}{\includegraphics{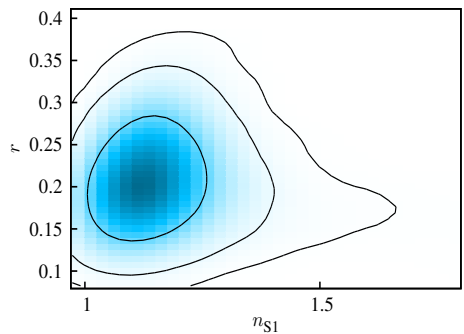}} 

\end{center}
\caption{\footnotesize\label{fig:ns1r} Two dimensional marginalized probabilities of $n_{\rm S1}$ and $r$. 
Note that the degeneracy between $r$ and the large scale spectral index $n_{\rm S1}$ is evident. A 
higher tensor amplitude always prefers higher blue tilt of scalar PPS at large scales since it
decreases the power at low-$\ell$ temperature anisotropy
and fits the Planck data better. Plots above represent the results where tensor spectral index fixed to $-r/8$.}
\end{figure*}

Since it is $n_{\rm S1}$ which causes the large scale suppression in power that fits both Planck and BICEP2 data better than power law, 
it is expected that $n_{\rm S1}$ and $r$ will have a degeneracy. We should mention that these are the two main parameters of the complete analysis since $n_{\rm S1}$ helps to fit 
Planck and $r$ helps to fit BICEP2 data and reconcile these two datasets. In Fig~\ref{fig:ns1r} we plot the 2d marginalized contours of 
$n_{\rm S1}-r$. We find that the degeneracy between these two parameters indicate that a higher $r$ is only allowed to fit BICEP2 data when we allow
more blue tilt in scalar PPS.

\begin{figure*}[!htb]
\vskip -40pt

\begin{center} 
\resizebox{210pt}{160pt}{\includegraphics{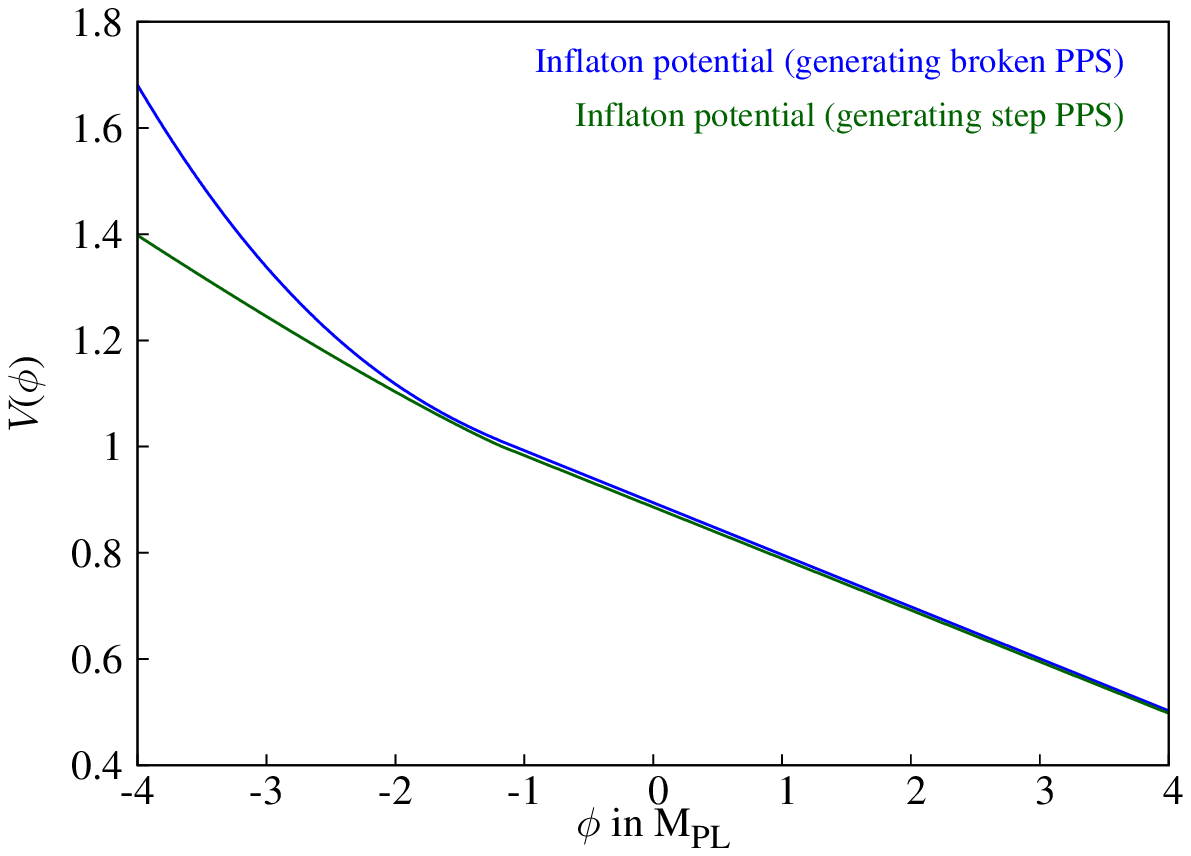}} 
\resizebox{210pt}{160pt}{\includegraphics{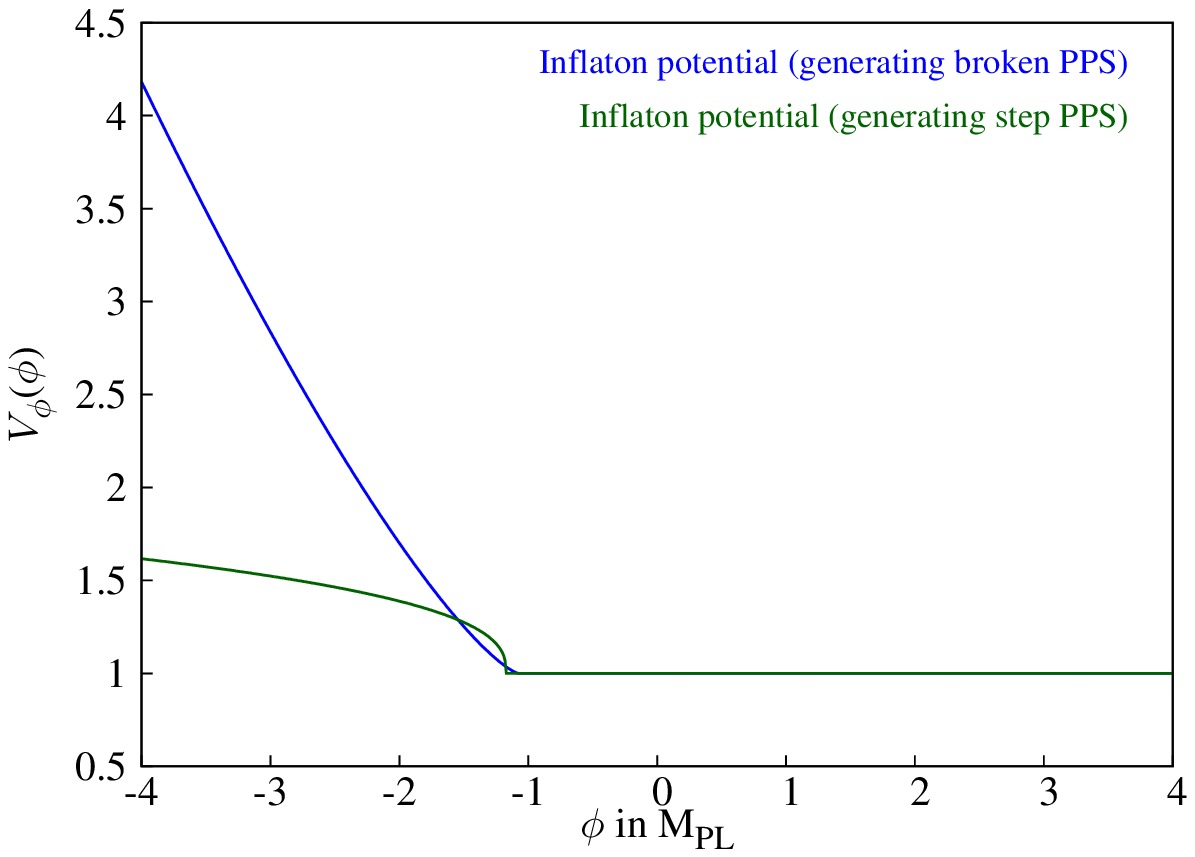}} 
\resizebox{210pt}{160pt}{\includegraphics{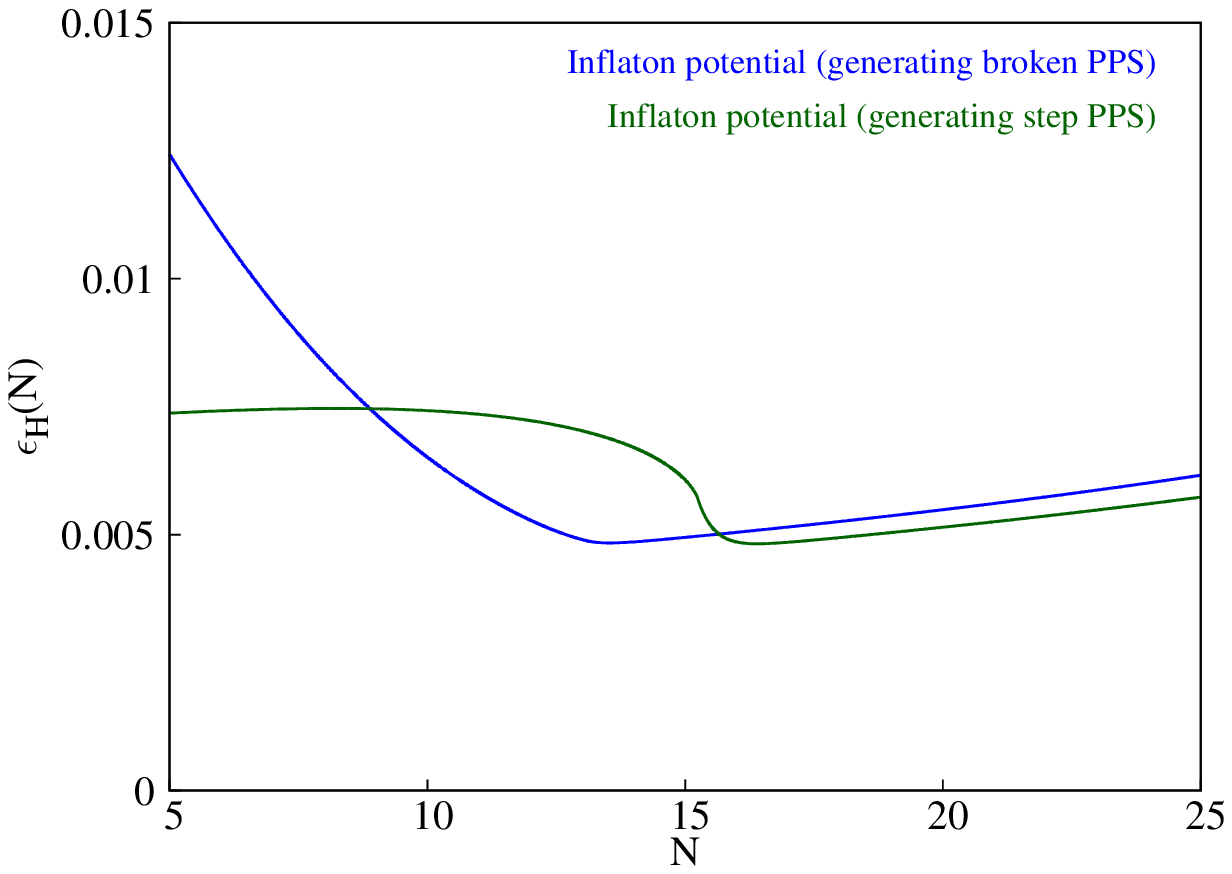}} 
\resizebox{210pt}{160pt}{\includegraphics{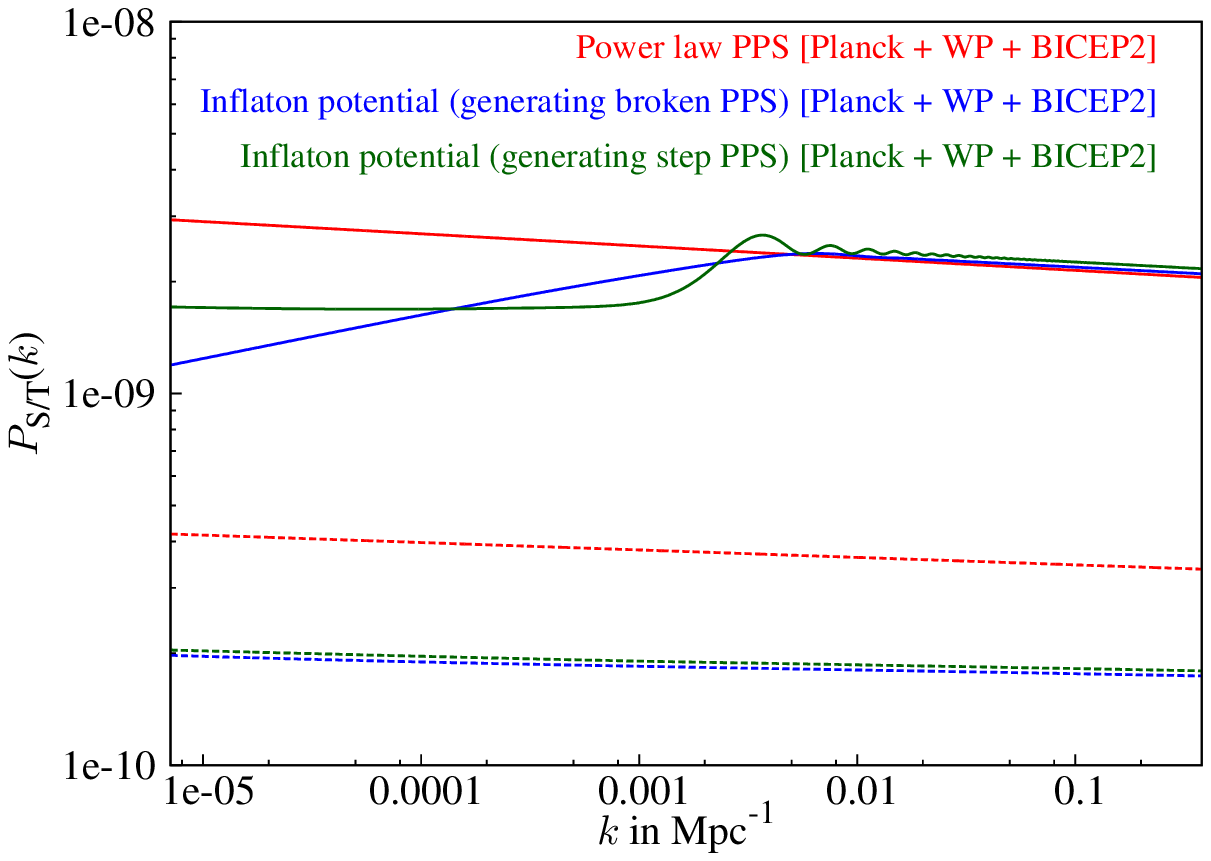}} 
\resizebox{210pt}{160pt}{\includegraphics{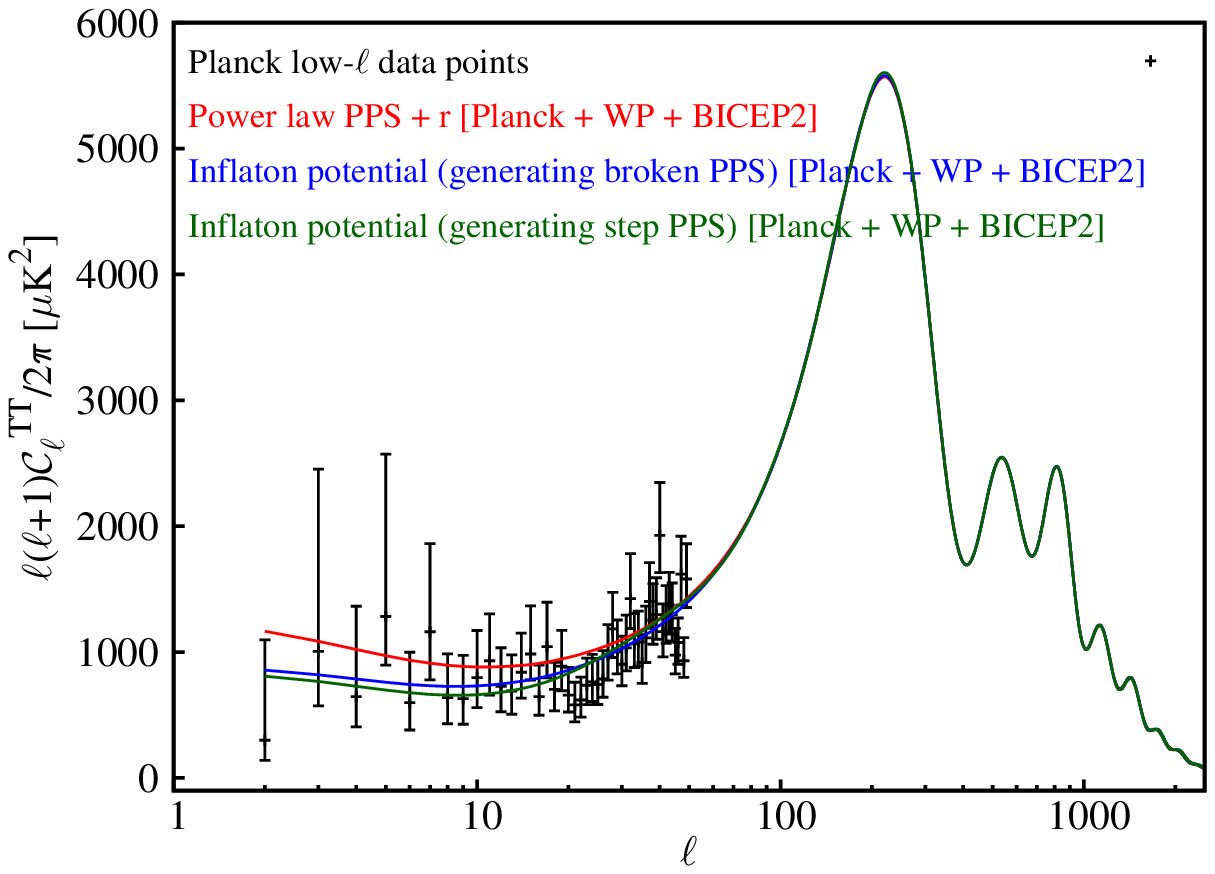}} 
\resizebox{210pt}{160pt}{\includegraphics{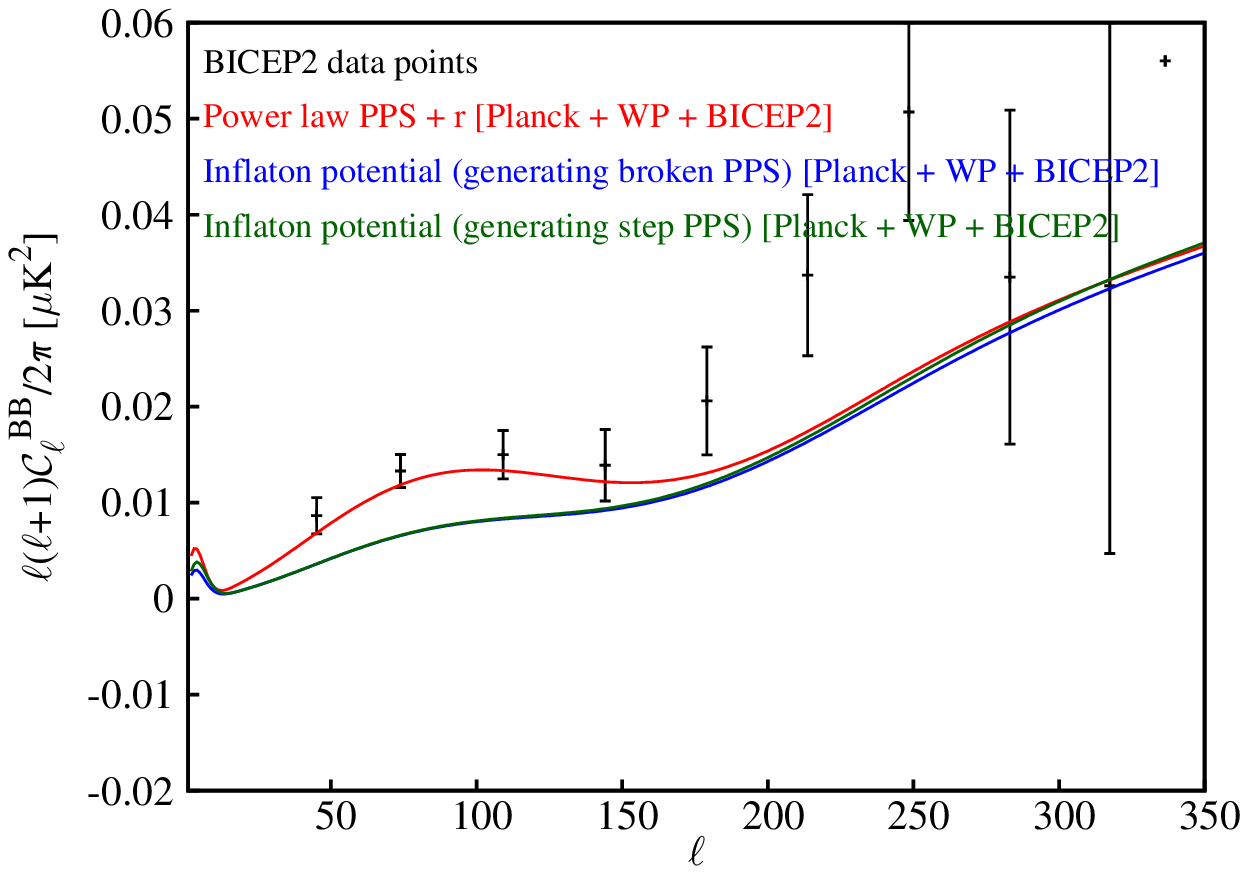}}

\end{center}
\caption{\footnotesize\label{fig:theory}The best fit results from the inflaton potential described in Eq.~\ref{eq:potential}. 
In all the plots the curves in blue represent the best fit result from the region of potential parameter space where we are able to get a broken type scalar PPS and the plots in 
green represent the best fit from the region providing a step type scalar PPS. Top Left and right : The best fit potentials and their derivatives (normalized at the transition)
are plotted. Middle left : The first slow roll parameter $\epsilon_{\rm H}=-\dot{H}/H^2$ for the two cases plotted. Note that the inflaton first rolls 
for around 15 e-folds before the break and then starts rolling slowly. Middle right : The best fit scalar (solid) and the tensor (dashed) PPS are plotted. The power law 
best fit are plotted in red. Note that the inflationary scalar PPS in both the cases matches power law at small scales but drops in power at
large scales reproducing the broken and Tanh scalar PPS discussed before. The tensor power spectrum, however, is lower than the power law best fit. 
Bottom left : The best fit $\cl^{\rm TT}$ are plotted along with the Planck low-$\ell$ data (in black) and the power law (in red). It is clear from this 
plot that a broken type or step type model is able to fit the Planck data better than the power law due to the suppression. Bottom right : The 
best fit $\cl^{\rm BB}$ along with BICEP2 data (in black) and the power law best fit (in red) are plotted. Due to the smaller tensor-to-scalar ratio 
$r\sim0.07-0.1$ the particular inflationary models discussed here are not able to fit the BICEP2 data similar to power law where $r$ is a 
free parameter.}
\end{figure*}

Having described the phenomenological models of the PPS we shall now present the results obtained from the inflaton potential. 
For this model we solve the background and the perturbation equations using {\tt BINGO}~\cite{Hazra:2012yn} without any approximations. 
We search for the best fit in the vicinity of the $\zeta\sim1.2$ (to generate a Tanh step like scalar PPS) and $\zeta\sim2$ 
(to generate a broken scalar PPS) using Powell's BOBYQA minimization algorithm. We
should mention that along with $\zeta$ we have allowed the other inflation potential parameters and background 
cosmological parameter and foreground nuisance
parameters to vary as well. The best fit results are plotted in Fig.~\ref{fig:theory}. In blue, we plot the results that closely 
resemble the our broken scalar PPS model and in green we plot the results that resemble the Tanh step model. The top panel
plots contain the best fit potential $V(\phi)$ (left) and its derivative $V_\phi(\phi)$ (right). The middle left panel contain 
the first slow roll parameter $\epsilon_{\rm H}=-\dot{H}/H^2$ (left). The best fit scalar (solid curves) and the tensor PPS (dashed 
curves) are plotted in the middle right
\clearpage
\noindent panel along with the power law best fit for Planck + WP + BICEP2 as a reference. Note that the 
tensor power spectra in these two models are nearly half in magnitude compared to the best fit tensor PPS for power law model. 
Due to the suppression in scalar power, these inflationary PPS can fit the Planck data better than power law (bottom left panel)
but fails to address the BICEP2 data (bottom right) due to less tensor power compared to the best fit values. 
As a result we find improvement in fit from Planck + WP compared to power
law but overall fit gets worse by ${\cal O}(3)$. However this model shows a balance between Planck and BICEP2 data combination. 
We are presently trying to build models that can generate the large scale suppression along with a tensor PPS with 
high amplitude to fit the BB data from BICEP2. Moreover, we should mention, since BICEP2 and KECK cross power spectra suggest that
the peak of the B-mode a bit suppressed~\cite{Ade:2013uln}, 
the potentials that we have used in this paper will address the the cross spectra better than BICEP2 data alone if the results persist.

\section{Conclusions}\label{sec:conclusions}

In this paper we show that power-law form of the primordial spectrum (scalar PPS) cannot fit properly Planck temperature and BICEP2 
B-Mode polarization data simultaneously. In fact scalar power-law form of PPS is disfavored at more than $3\sigma$ in comparison 
with the kink model we have studied. There have been hints in the Planck temperature data alone (along with WMAP low $\ell$ E-mode 
polarization data) that a broken form of the PPS can fit the data pretty well as we have discussed it in details 
in~\cite{Hazra:2013nca}. However, due to cosmic variance, it was not possible to favor this phenomenological model to power-law 
form of the primordial spectrum with high confidence using temperature data alone. Using BICEP2 B-mode polarization data this 
degeneracy seems to be broken now and we have estimated that the power-law form of PPS is ruled out at more than $3\sigma$ CL, 
assuming inflationary consistency relation. It is indeed interesting to see that a simple broken form of the PPS can indeed fit both
data pretty well unlike power-law form of the primordial spectrum. This is evident by looking at the overlap of the confidence contours
in Fig.~\ref{fig:icon1d2d} fitting Planck data alone and fitting combination of Planck and BICEP2 data for two 
cases of power-law and broken form of 
the PPS. Ruling out the power-law form of the scalar primordial spectrum is an important result since it is one of the main assumptions of the 
concordance model of cosmology and at the same time there are many inflationary scenarios that result to power-law form of the primordial
spectrum. If confirmed, the detection of CMB B-mode polarization is a major discovery and opens windows. An example is that these 
new results seems to be quite severe for the existing standard model of cosmology for investigation of fine structure of inflation.
We have also shown that a Tanh step form of the PPS, again what we proposed and discussed in~\cite{Hazra:2013nca}, can also fit properly
both Planck and BICEP2 data simultaneously. These results reflects the fact that era of the Vanilla concordance model of cosmology is 
near to end (if the observational data persist) and we need more flavour to explain our observable Universe. We may need about 3 
additional parameters to express the initial perturbation such as the value of $r$ (tensor-to-scalar ratio), an additional spectral
index for the low-$k$ wavenumbers and a $k_{\rm b}$, the wavenumber where the break/transition can occur. We should note that there is an 
advantage for the broken or step forms of the scalar PPS to assumption of running of the scalar spectral index since 
by assuming running we limit ourself to a particular shape and thereby allow less flexibility. It is interesting to note 
that by assuming running there is one less degree of freedom in comparison to the broken PPS, but the power-law form of the PPS is ruled out  
with higher confidence by assuming the broken PPS. We should note here that the running of the spectral index, $d n_{\rm S}/d\ln k$ 
needs to have a large negative value $\sim -0.02$~\cite{Ade:2014xna} in order to fit both Planck and BICEP2. This large negative running
introduces large suppression in scalar PPS at small scales which needs to be balanced by including another parameter, for example, running of running or 
neutrinos. By just considering the running of the scalar spectral index one can get about 6.5 improvement in the $\chi^2$ in
comparison to power law, while by considering a broken power spectrum we can get an improvement of about 12-13. It is in fact double 
improvement in the $\Delta \chi^2$ by having only one more extra parameter. Our result signifies the importance of the one additional 
degree of freedom in the broken scalar PPS (compared to running), {\it i.e.} the position of the break $k_{\rm b}$. This shows that the 
assumption of the running of the spectral index may not suffice to explain the data properly. Our results also indicate that relaxing
the inflationary consistency relation can help the power-law form of the PPS to be less inconsistent to the data (but still considerably 
inconsistent) but it would not improve the fit much for the broken or step forms of the PPS. This is a good news since it seems by assuming 
these simple non-power-law forms of the PPS, there will not be any tension between various CMB data and we can still hold on the theoretically
important inflationary consistency relation. 

From the theoretical perspective, once we fix an inflationary model potential, we fix the amplitude and tilt of the scalar and 
tensor power spectra simultaneously. There is no more freedom to change $r$ within a model. Our preliminary results presented in this work show
that within an inflationary scenario, previously discussed in~\cite{Bousso:2013uia} we can get the scalar PPS that can resemble the 
broken and the step like scalar PPS discussed in this paper, but it can not generate large tensor component that can address BICEP2 
data well. If observational constraints for $r$ changes to values close to $~0.1$ rather than the current central value of $0.2$, 
there seems to be more space for inflationary scenarios to explain all data simultaneously.
Thus, the BICEP2 discovery of primordial gravitational waves, while confirming the general 
observational prediction~\cite{S79} of the of the early Universe scenario with the de Sitter (inflationary)
stage preceding the hot radiation dominated stage, shows that the inflationary stage is not so simple and may not be described by a
one-parametric model. 
We focus on inflationary model (Whipped Inflation) building in a separate paper~\cite{Hazra:2014jka} wherein we discuss that using 
canonical scalar fields, generation of large tensors, suppression in scalar power at large scales and at the same time low level of 
non-Gaussianity is achievable. 
We should also mention that expansion beyond linear order term in the slow roll part of the potential in~\cite{Bousso:2013uia} 
can also help in generating a large tensor amplitude and hence can address BICEP2 data better than the potential we have used in this paper~\cite{Bousso:2014jca}.  
We also wish to perform a complete parameter estimation for the inflationary models we discussed, where we expect the B-mode polarization data
from POLARBEAR~\cite{Ade:2014afa} will help us in providing tighter constraints.

Just before finishing this paper it is important to address about an alternative approach which has been proposed to reconcile Planck and 
BICEP2 data within the frame of the power-law form of the primordial spectrum by assuming extra massive sterile neutrino. Soon after release of 
the Planck data it was realized that by assuming an additional massive sterile neutrino, the scale invariant form of the primordial spectrum ($n_s=1$) 
can be consistent to the data while the Hubble parameter should have higher values (which could make it even more consistent to the local Universe 
estimations of $H_0$)~\cite{Ade:2013zuv}. After release of BICEP2 data one could guess that this model may work well to fit both Planck and BICEP2 as
well. In~\cite{Zhang:2014dxk,Dvorkin:2014lea} it was shown that such model can indeed fit the combination of Planck and BICEP2 data reasonably well 
while staying within the context of the power-law form of the primordial spectrum. We should note that this model has limited flexibilities to suppress 
low $\ell$ scalar multipoles and also to describe the fine structure of the temperature spectrum. We would have soon E-Mode polarisation data from Planck 
(where the effect of broken form of the primordial spectrum and having massive sterile neutrino would be different on the data) that will help us to 
differentiate between these two main alternatives with high confidence.

We should mention here that there have been some publications~\cite{Contaldi:2014zua,Miranda:2014wga,Abazajian:2014tqa,Ashoorioon:2014nta} in last 
few days after release of the BICEP2 data that we may share some of the results, however we should 
emphasize here that this paper is in fact a straightforward extension of the ~\cite{Hazra:2013nca} considering BICEP2 data.


\section*{Acknowledgments}
D.K.H. and A.S. wishes to acknowledge support from the Korea Ministry of Education, Science and Technology, Gyeongsangbuk-Do and Pohang 
City for Independent Junior Research Groups at the Asia Pacific Center for Theoretical Physics. We also acknowledge the use of 
publicly available {\tt CAMB} and {\tt COSMOMC} in our analysis.
The authors would like to thank Antony Lewis for providing us the new {\tt COSMOMC} package that takes into account the recent 
BICEP2 data. We acknowledge the use of WMAP-9 data 
and from Legacy Archive for Microwave Background Data Analysis (LAMBDA)~\cite{lambdasite}, Planck 
data and likelihood from Planck Legacy Archive (PLA)~\cite{PLA} and BICEP2 data from~\cite{biceprepo}. A.S. would like to acknowledge
the support of the National Research Foundation of Korea (NRF-2013R1A1A2013795). A.A.S. was also 
partially supported by the grant RFBR 14-02-00894. 



\begin{thebibliography}{99}
\bibitem{Ade:2013kta} 
  P.~A.~R.~Ade {\it et al.}  [Planck Collaboration],
  arXiv:1303.5075 [astro-ph.CO].
  
\bibitem{Ade:2014gua} 
  P.~A.~R.~Ade {\it et al.}  [BICEP2 Collaboration],
  arXiv:1403.4302 [astro-ph.CO].
\bibitem{Ade:2014xna} 
  P.~A.~R.~Ade {\it et al.}  [BICEP2 Collaboration],
  arXiv:1403.3985 [astro-ph.CO].
 \bibitem{Hazra:2013nca} 
  D.~K.~Hazra, A.~Shafieloo and G.~F.~Smoot,
   JCAP {\bf 1312}, 035 (2013)
  arXiv:1310.3038 [astro-ph.CO].

\bibitem{Bousso:2013uia} 
  R.~Bousso, D.~Harlow and L.~Senatore,
  arXiv:1309.4060 [hep-th].
\bibitem{Peiris:2003ff} 
  H.~V.~Peiris {\it et al.}  [WMAP Collaboration],
  Astrophys.\ J.\ Suppl.\  {\bf 148}, 213 (2003)
  [astro-ph/0302225].
\bibitem{Hinshaw:2012fq}
  G.~Hinshaw, D.~Larson, E.~Komatsu, D.~N.~Spergel, C.~L.~Bennett, J.~Dunkley, M.~R.~Nolta and M.~Halpern {\it et al.},
  arXiv:1212.5226 [astro-ph.CO].
  
\bibitem{reconstruction-all}
  S.~Hannestad,
  Phys.\ Rev.\ D {\bf 63} (2001) 043009
  [astro-ph/0009296];
  M.~Tegmark and M.~Zaldarriaga,
  Phys.\ Rev.\ D {\bf 66} (2002) 103508
  [astro-ph/0207047];
%
%
  S.~L.~Bridle, A.~M.~Lewis, J.~Weller and G.~Efstathiou,
  Mon.\ Not.\ Roy.\ Astron.\ Soc.\  {\bf 342} (2003) L72
  [astro-ph/0302306];
  A.~Shafieloo and T.~Souradeep,
  Phys.\ Rev.\ D {\bf 70} (2004) 043523
  [astro-ph/0312174];
  P.~Mukherjee and Y.~Wang,
  Astrophys.\ J.\  {\bf 599} (2003) 1
  [astro-ph/0303211];
%
%
%
%
%
  D.~Tocchini-Valentini, Y.~Hoffman and J.~Silk,
  Mon.\ Not.\ Roy.\ Astron.\ Soc.\  {\bf 367} (2006) 1095
  [astro-ph/0509478];
%
%
  N.~Kogo, M.~Sasaki and J.~'i.~Yokoyama,
  Prog.\ Theor.\ Phys.\  {\bf 114} (2005) 555
  [astro-ph/0504471];
%
%
  S.~M.~Leach,
  Mon.\ Not.\ Roy.\ Astron.\ Soc.\  {\bf 372} (2006) 646
  [astro-ph/0506390];
  A.~Shafieloo and T.~Souradeep,
  Phys.\ Rev.\ D {\bf 78} (2008) 023511
  [arXiv:0709.1944 [astro-ph]];
%
  P.~Paykari and A.~H.~Jaffe,
  Astrophys.\ J.\  {\bf 711} (2010) 1
  [arXiv:0902.4399 [astro-ph.CO]];
%
%
  G.~Nicholson and C.~R.~Contaldi,
  JCAP {\bf 0907}, 011 (2009)
  [arXiv:0903.1106 [astro-ph.CO]];
%
%
%
%
%
  C.~Gauthier and M.~Bucher,
  JCAP {\bf 1210}, 050 (2012)
  [arXiv:1209.2147 [astro-ph.CO]];
%
 R.~Hlozek, J.~Dunkley, G.~Addison, J.~W.~Appel, J.~R.~Bond, C.~S.~Carvalho, S.~Das and M.~Devlin {\it et al.},
  Astrophys.\ J.\  {\bf 749} (2012) 90
  [arXiv:1105.4887 [astro-ph.CO]];
%
  J.~A.~Vazquez, M.~Bridges, M.~P.~Hobson and A.~N.~Lasenby,
  JCAP {\bf 1206}, 006 (2012)
  [arXiv:1203.1252 [astro-ph.CO]];
    D.~K.~Hazra, A.~Shafieloo and T.~Souradeep,
  JCAP {\bf 1307}, 031 (2013)
  [arXiv:1303.4143 [astro-ph.CO]];
    D.~K.~Hazra, A.~Shafieloo and T.~Souradeep,
  Phys.\ Rev.\ D {\bf 87}, 123528 (2013)
  [arXiv:1303.5336 [astro-ph.CO]];
  P.~Hunt and S.~Sarkar,
  arXiv:1308.2317 [astro-ph.CO].  
  

  
\bibitem{Ade:2013uln} 
  P.~A.~R.~Ade {\it et al.}  [Planck Collaboration],
  arXiv:1303.5082 [astro-ph.CO].

  \bibitem{JSS08}

M.~Joy, V.~Sahni, A.~A.~Starobinsky, Phys. Rev. D {\bf 77}, 023514 (2008) [arXiv:0711.1585].

\bibitem{JSSS09}

M.~Joy, A.~Shafieloo, V.~Sahni, A.~A.~Starobinsky. JCAP {\bf 0906}, 028 (2009) [arXiv:0807.3334].
  
\bibitem{S92}

A.~A.~Starobinsky, JETP Lett. {\bf 55}, 489 (1992).

\bibitem{CPKL03}
  C.~R.~Contaldi, M.~Peloso, L.~Kofman and A.~D.~Linde,
  JCAP {\bf 0307}, 002 (2003)
  [astro-ph/0303636].
  
\bibitem{Linde:1998iw} 
  A.~D.~Linde,
  Phys.\ Rev.\ D {\bf 59}, 023503 (1999)
  [hep-ph/9807493].
\bibitem{Linde:1999wv} 
  A.~D.~Linde, M.~Sasaki and T.~Tanaka,
  Phys.\ Rev.\ D {\bf 59}, 123522 (1999)
  [astro-ph/9901135].
  

\bibitem{Sato:1980}
K.~Sato, Mon. Not. R. astr. Soc (1981) {\bf 195}, 467-479.   



\bibitem{S98}

A.~A.~Starobinsky, Gravit. Cosmol. {\bf 4}, Suppl., 88 (1998)
[arXiv:astro-ph/9811360].


\bibitem{features} 
  R.~Allahverdi, K.~Enqvist, J.~Garcia-Bellido and A.~Mazumdar,
  Phys.\ Rev.\ Lett.\  {\bf 97} (2006) 191304
  [hep-ph/0605035];
J.~Hamann, L.~Covi, A.~Melchiorri and A.~Slosar,
  Phys.\ Rev.\ D {\bf 76} (2007) 023503
  [astro-ph/0701380];
  R.~Allahverdi, K.~Enqvist, J.~Garcia-Bellido, A.~Jokinen and A.~Mazumdar,
  JCAP {\bf 0706} (2007) 019
  [hep-ph/0610134]; 
  R.~K.~Jain, P.~Chingangbam, J.~-O.~Gong, L.~Sriramkumar and T.~Souradeep,
  JCAP {\bf 0901} (2009) 009
  [arXiv:0809.3915 [astro-ph]];
M.~J.~Mortonson, C.~Dvorkin, H.~V.~Peiris and W.~Hu,
  Phys.\ Rev.\ D {\bf 79}, 103519 (2009)
  [arXiv:0903.4920 [astro-ph.CO]];
  R.~K.~Jain, P.~Chingangbam, L.~Sriramkumar and T.~Souradeep,
  Phys.\ Rev.\ D {\bf 82} (2010) 023509
  [arXiv:0904.2518 [astro-ph.CO]];
    D.~K.~Hazra,
  JCAP {\bf 1303}, 003 (2013)
  [arXiv:1210.7170 [astro-ph.CO]];
   E.~Dudas, N.~Kitazawa, S.~P.~Patil and A.~Sagnotti,
  JCAP {\bf 1205}, 012 (2012)
  [arXiv:1202.6630 [hep-th]];
    N.~Kitazawa and A.~Sagnotti,
  arXiv:1402.1418 [hep-th].
\bibitem{Hazra:2010ve} 
  D.~K.~Hazra, M.~Aich, R.~K.~Jain, L.~Sriramkumar and T.~Souradeep,
  JCAP {\bf 1010}, 008 (2010)
  [arXiv:1005.2175 [astro-ph.CO]].


%
%
%
%
%
%
%
%
%
%
%
%
%
%
%
%
%
%
%
%
\bibitem{Hazra:2012yn} 
  D.~K.~Hazra, L.~Sriramkumar and J.~Martin,
  JCAP {\bf 1305}, 026 (2013)
  [arXiv:1201.0926 [astro-ph.CO]].

 \bibitem{cambsite}
See {\tt http://camb.info/.}
\bibitem{Lewis:1999bs}
  A.~Lewis, A.~Challinor and A.~Lasenby,
  Astrophys.\ J.\  {\bf 538} (2000) 473
  [astro-ph/9911177].
%
\bibitem{cosmomcsite}
See {\tt http://cosmologist.info/cosmomc/.}


\bibitem{Lewis:2002ah}
  A.~Lewis and S.~Bridle,
  Phys.\ Rev.\ D {\bf 66} (2002) 103511
  [astro-ph/0205436].
    



\bibitem{powell}
  M.~J.~D.~Powell, Cambridge NA Report NA2009/06, University of Cambridge, Cambridge (2009).
  
  \bibitem{S79}

A.~A.~Starobinsky, JETP Lett. 30, 682 (1979).
\bibitem{Hazra:2014jka} 
  D.~K.~Hazra, A.~Shafieloo, G.~F.~Smoot and A.~A.~Starobinsky,
  arXiv:1404.0360 [astro-ph.CO].
\bibitem{Bousso:2014jca} 
  R.~Bousso, D.~Harlow and L.~Senatore,
  arXiv:1404.2278 [astro-ph.CO].

\bibitem{Ade:2014afa} 
  P.~A.~R.~Ade {\it et al.}  [ The POLARBEAR Collaboration],
  arXiv:1403.2369 [astro-ph.CO].
  
\bibitem{Ade:2013zuv} 
  P.~A.~R.~Ade {\it et al.}  [Planck Collaboration],
  arXiv:1303.5076 [astro-ph.CO].


\bibitem{Zhang:2014dxk} 
  J.~-F.~Zhang, Y.~-H.~Li and X.~Zhang,
  arXiv:1403.7028 [astro-ph.CO].


\bibitem{Dvorkin:2014lea} 
  C.~Dvorkin, M.~Wyman, D.~H.~Rudd and W.~Hu,
  arXiv:1403.8049 [astro-ph.CO].


\bibitem{Contaldi:2014zua} 
  C.~R.~Contaldi, M.~Peloso and L.~Sorbo,
  arXiv:1403.4596 [astro-ph.CO].
\bibitem{Miranda:2014wga} 
  V.~íc.~Miranda, W.~Hu and P.~Adshead,
  arXiv:1403.5231 [astro-ph.CO].
\bibitem{Abazajian:2014tqa} 
  K.~N.~Abazajian, G.~Aslanyan, R.~Easther and L.~C.~Price,
  arXiv:1403.5922 [astro-ph.CO].
\bibitem{Ashoorioon:2014nta} 
  A.~Ashoorioon, K.~Dimopoulos, M.~M.~Sheikh-Jabbari and G.~Shiu,
  arXiv:1403.6099 [hep-th].



\bibitem{lambdasite}  
See {\tt http://lambda.gsfc.nasa.gov/product/map/dr3/m$\_$products.cfm.}

\bibitem{PLA}  
See \\
{\tt http://www.sciops.esa.int/index.php?project=planck$\&$page=Planck$\_$Legacy$\_$Archive.}

\bibitem{biceprepo}  
See {\tt http://bicepkeck.org/\#data$\_$products.}

  
%
%
%
%
%
%
%
%
%
%
  


%
%
%
%
%
%
%
%
%
%
%
%
%
%
%
%
%
%
%
%
%
%
%
%
%
%
%
%
%
%
%
%
%
%
%
%
%
%
%
%
%
%
%
%
%
%
%
%
%
%
%
%
%
%
%
%
%
%
%
%
%
%
%
%
%
%
%
%
%
%
%
%
%
%
%
%
%
%
%
%
%
%
%
%
%
%
%
%
%
%
%


  
\end{thebibliography}
\end{document}